\documentclass[a4paper, twocolumn, 10pt, accepted=2024-10-15]{quantumarticle}
\pdfoutput=1 %
\usepackage[utf8]{inputenc}
\usepackage[english]{babel}
\usepackage{amsmath}
\usepackage[binary-units=true]{siunitx}
\usepackage{booktabs}
\usepackage{combelow}
\usepackage{csquotes}
\usepackage[colorlinks,citecolor=blue,bookmarks=true]{hyperref}
\usepackage[nameinlink]{cleveref}  
 \usepackage{amsmath}
 \usepackage{amssymb}
 \usepackage{amsthm}
 \usepackage{amsfonts}
 \usepackage{subcaption}

\newtheorem{example}{Example}[section]

\usepackage[numbers, sort&compress]{natbib} %
\newcommand{\LO}{{\fontfamily{qcr}\selectfont LightsOut}}

\begin{document}

\title{Decoding quantum color codes with MaxSAT}

\author{Lucas Berent}
\email{lucas.berent@tum.de}
\orcid{0000-0002-2973-1689}
\affiliation{Technical University of Munich, Germany}

\author{Lukas Burgholzer}
\email{lukas.burgholzer@jku.at}
\orcid{0000-0003-4699-1316}
\affiliation{Johannes Kepler University Linz, Austria}

\author{Peter-Jan H.~S. Derks}
\email{peter-janderks@hotmail.com}
\affiliation{Freie Universit{\"a}t Berlin, Germany}
\orcid{0000-0002-9197-1309}

\author{Jens Eisert}
\email{jense@zedat.fu-berlin.de}
\orcid{0000-0003-3033-1292}
\affiliation{Freie Universit{\"a}t Berlin, Germany}
\affiliation{Helmholtz-Zentrum Berlin f{\"u}r Materialien und Energie, Germany}

\author{Robert Wille}
\email{robert.wille@tum.de}
\orcid{0000-0002-4993-7860}
\affiliation{Technical University of Munich, Germany}
\affiliation{Software Competence Center Hagenberg GmbH (SCCH), Austria}

\maketitle

\begin{abstract}
In classical computing, error-correcting codes are well established and are ubiquitous both in theory and practical applications. For quantum computing, error correction is essential as well, but harder to realize, coming along with substantial resource overheads and being concomitant with the need for substantial classical computing. Quantum error-correcting codes play a central role on the avenue towards fault-tolerant quantum computation beyond presumed near-term applications. Among those, color codes constitute a particularly important class of quantum codes that have gained interest in recent years due to favourable properties over other codes. As in classical computing, \emph{decoding} is the problem of inferring an operation to restore an uncorrupted state from a corrupted one and is central in the development of fault-tolerant quantum devices. In this work, we show how the decoding problem for color codes can be reduced to a slight variation of the well-known \LO{} puzzle. We propose a novel decoder for quantum color codes using a formulation as a MaxSAT problem based on this analogy. Furthermore, we optimize the MaxSAT construction and show numerically that the decoding performance of the proposed decoder achieves state-of-the-art decoding performance on color codes. The implementation of the decoder, as well as tools to automatically conduct numerical experiments, are publicly available as part of the \emph{Munich Quantum Toolkit} (MQT) at~\url{https://github.com/cda-tum/mqt-qecc}. 
\end{abstract}

\section{Introduction}
In classical computing, error-correcting codes are ubiquitous
both in theory and practical applications. The main idea is to add redundancy to data that needs to be protected from errors in order to tolerate noise and enable the correction of errors. A famous class of codes are so-called linear codes, which are defined as a vector space that allows to encode $k$ \emph{logical} bits (that hold the information) into $n$ \emph{physical} bits, where $n>k$. The simplest form of an error-correcting code is the $n$-repetition code, which encodes $0\mapsto 000$ and $1 \mapsto 111$ (for $n=3$). In quantum computing~\cite{nielsen2002quantum}, a similar need for error-correction exists: 
qubits are fragile in nature and experience decoherence over time. Moreover, operations on qubits are imperfect and thus introduce additional errors. As a consequence, running a quantum algorithm consisting of thousands of operations on such a noisy machine produces results that are almost completely random. Thus, to realize large-scale quantum computers, it is necessary to protect quantum systems from inevitable noise occurring during computation.

To overcome this issue of noise, \emph{quantum error-correction} (QEC) and -- building upon this, allowing each step to be imperfect -- \emph{fault-tolerance} (FT) will ultimately be needed. The general idea---similar to classical coding theory---is to use \emph{quantum error-correcting codes} (QECCs) to encode quantum information by using additional, redundant information. However, due to the laws of quantum physics, for instance, the famous \emph{no-cloning theorem}, naive mechanisms such as copying quantum data are not possible. In order to conduct computations on encoded data, referred to as the \emph{logical information}, universal sets of noise-resilient quantum operations need to be designed, which is the goal of research in quantum fault-tolerance~\cite{preskill1998reliable, shor1996fault}. 

The general idea behind QEC is to conduct specific measurements on the (possibly corrupted) encoded information without destroying the encoded logical information.
This process allows inferring whether an error occurred and approximately ``where'' this error occurred while acquiring no information whatsoever about
the encoded logical information. Given this classical measurement information, the aim is to derive a \emph{recovery operation} that, when applied to the corrupted state, restores the encoded information to an uncorrupted state. The process of inferring such a recovery operation is called \emph{decoding} and is highly non-trivial---as in classical coding theory. In fact,
notions of decoding are moving much to the centre of the discussion of
quantum error correction. This is for good reason, as the classical
software needs to keep up with the time scales of the
quantum noise and has to offer correction steps in real time while the noise is acting on the system. For this reason, decoders must not only have high decoding performance but also be fast.

An important class of quantum codes is constituted by \emph{Calderbank-Shor-Steane} (CSS) codes~\cite{calderbank1996good, steane1996error}, which can be defined as a suitable combination of two classical linear codes. Out of CSS codes, \emph{planar color codes}~\cite{bombin2006topological,kesselring2022anyon, kubica2018abcs,thomsen2022low} form an essential class. They are defined on a two-dimensional lattice that is required to be three-valent and three-colourable with respect to its faces. There are various properties such as lower resource requirements, or the realization of logical operations that render color codes favourable over the currently widely researched \emph{surface code}~\cite{kitaev2003fault, krinner2022realizing, google2023suppressing}. Moreover, there have been first advances of physical implementations of FT-quantum computation with small instances of color codes~\cite{postler2022demonstration, ryan2021realization}. 
One of the central open problems towards FT is scalable, accurate and fast decoding. After all, in any real implementation, decoding has to be faster than the quantum noise can compromise the coherence of the quantum information stored.

In this work, we show that the decoding problem can be formulated as a slight variation of the well-known \LO{}-puzzle. \LO{} is a famous combinatorial puzzle and has various connections to linear algebra and graph theory. Based on this analogy, we propose a MaxSAT encoding that allows to determine minimal solutions to the underlying problem. By this, we aim to bridge the gap between powerful classical constraint satisfaction solving approaches and quantum computing and thereby inspire new research in this direction.

\begin{figure*}
    \centering    
    \includegraphics[width=0.9\textwidth]{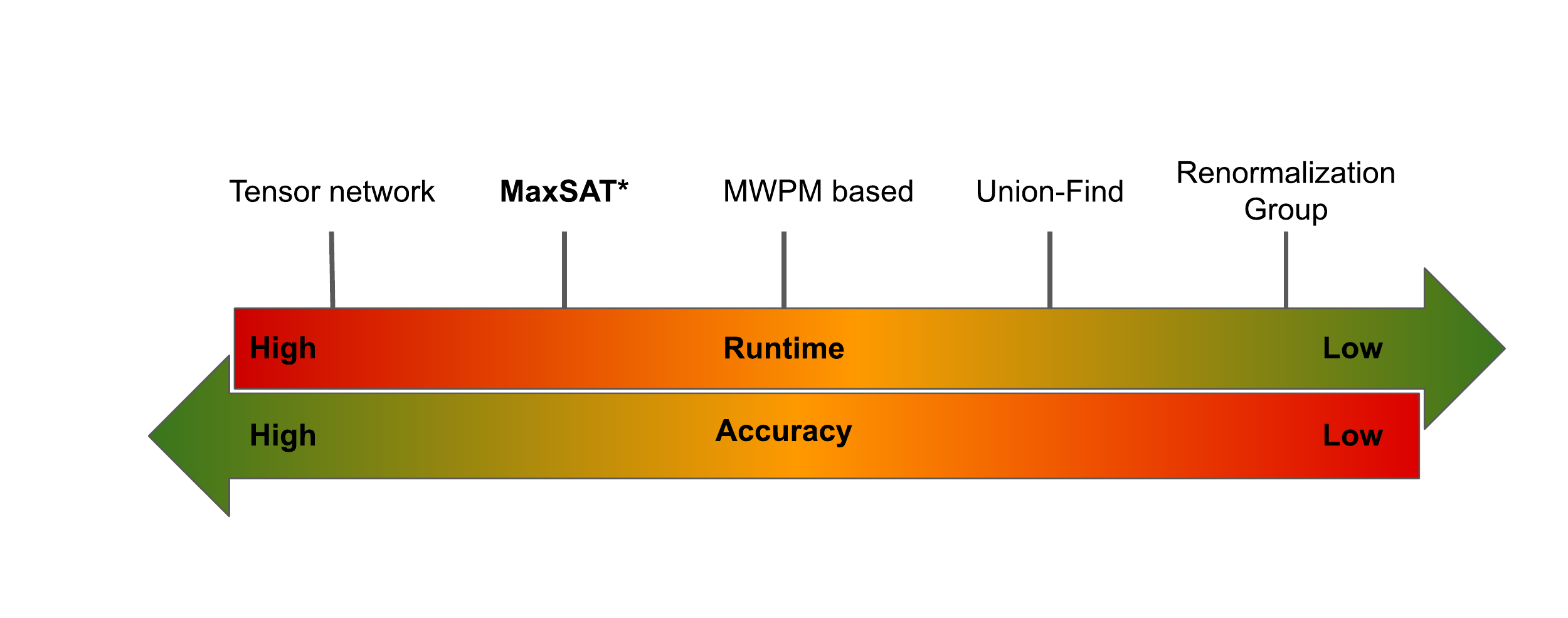}
    \caption{Abstract comparison of the decoding accuracy and runtime of several color code decoders. This figure is inspired by a figure comparing surface code decoders in Ref.~\cite{NewmanTalk}.}
    \label{fig:decoder_comparison}
\end{figure*}

Numerical evaluations of the resulting MaxSAT decoder for color codes show that it achieves near-optimal decoding performance for the considered noise model (bit-flip errors with noiseless syndrome measurements). 
Our decoder outperforms all color code decoders in terms of accuracy, except the tensor network decoder~\cite{chubb2021general}, which it outperforms in runtime below the threshold. \Cref{fig:decoder_comparison} shows a classification of color code decoders. To achieve these results, we improved the performance as well as the re-usability of the proposed encoding by carefully optimizing all constraints. Moreover, we compared the runtime of different MaxSAT solvers for the proposed satisfiability encoding. The results show that the implementation with Z3 outperforms other solvers.

Finally, both the implementation of the proposed decoder and means to conduct numerical simulations are provided as an open-source Python package as part of the \emph{Munich Quantum Toolkit} (MQT)~\cite{10646543} at~\url{https://github.com/cda-tum/mqt-qecc}. 

The remainder of this work is structured as follows. In ~\Cref{sec:background}, a high-level overview on quantum color codes is given to provide some background on quantum CSS codes. Then, in~\Cref{sec:overview} we give an overview of the main problem and discuss prior work. In~\Cref{sec:proposed-method}, the \LO{} puzzle and the analogy to color code decoding are discussed. The corresponding MaxSAT formulation is presented in~\Cref{sec:maxsat}. Then, numerical experiments and results are given in~\Cref{sec:experiments}. Finally, a summary and brief outlook on future work is presented in~\Cref{sec:conclusion}.

\section{Background}\label{sec:background} %
In this section, a brief introduction to quantum error correction and color codes is given.  For the sake of simplicity, the fundamental notions are explained in a high-level manner and we refer the reader to introductions on quantum error-correction and color codes~\cite{nielsen2002quantum, roffe2019quantum, kesselring2022anyon,kesselring2018boundaries,thomsen2022low, kubica2018abcs} for more details.

A classical linear code $\mathcal{C}$ that encodes $k$ logical bits into $n$ physical bits with distance $d$, written as an $[n,k,d]$-code, is defined as a $k$-dimensional vector space
\[
    \mathcal{C} = \{c \in \mathbb{F}_2^n: Hc=0\},
\]
i.e., the kernel of a matrix $H\in \mathbb{F}_{2}^{(n-k)\times n}$, called the \emph{parity-check matrix} of $\mathcal{C}$. Intuitively, the rows of $H$ correspond to parity checks and a vector $c\in \mathbb{F}_{2}^n$ is a \emph{codeword} iff all checks are satisfied, i.e., all checks are equal to $0$. The \emph{distance} $d$ of the code is the minimum Hamming weight of a non-zero codeword. In principle, a code with distance $d$ can correct up to $(d-1)/2$ errors.

\begin{example}\label{ex:hammingc}
Consider the parity-check matrix $$H = \left(\begin{matrix}
				1 &0 &0 &1 &0 &1 &1 \\
				0 &1 &0 &1 &1 &0 &1 \\
				0 &0 &1 &0 &1 &1 &1\\
			    \end{matrix}\right),$$ 
       which defines a $[7,4,3]$-code. The vector $x = (1,0,0,0,0,0)$ is not a codeword, since the first check $x_0 + x_3 + x_5 + x_6 \overset{?}{=} 0$ is violated, i.e., $H\cdot x = (1,0,0) \neq (0,0,0)$. 
\end{example}

A quantum CSS code is defined as a combination of two classical linear codes with parity-check matrices $H_X\in \mathbb{F}_{2}^{r_X\times n}$ and $H_Z\in \mathbb{F}_{2}^{r_Z \times n}$ that fulfill the orthogonality condition 
\[H_ZH_X^T = 0 \, (\text{or equivalently } H_XH_Z^T=0).\] 
In quantum error correction two types of errors need to be corrected: phase and bit flip errors, denoted as $Z$ and $X$.
These errors can be detected using two sets of checks, $X$ and $Z$ checks, which correspond to a certain type of measurement of the encoded state. 
Analogously to the classical case, these checks can be understood as parity measurements of subsets of qubits and thus be written as parity check matrices.
The indices of the 1 entries in the rows of $H_X$ $(H_Z)$ prescribe which qubits to include in $X$ $(Z)$ checks.

Planar color codes form a subclass of CSS codes that have the same $H_X$ and $H_Z$ checks (i.e., they are \emph{self-dual} CSS codes). 
Thus, both sets of checks can be treated analogously and, for the rest of this work, without loss of generality, only a single set of checks is considered, which is referred to as $H$.

\begin{figure}
    \centering
    \includegraphics[width=0.5\textwidth]{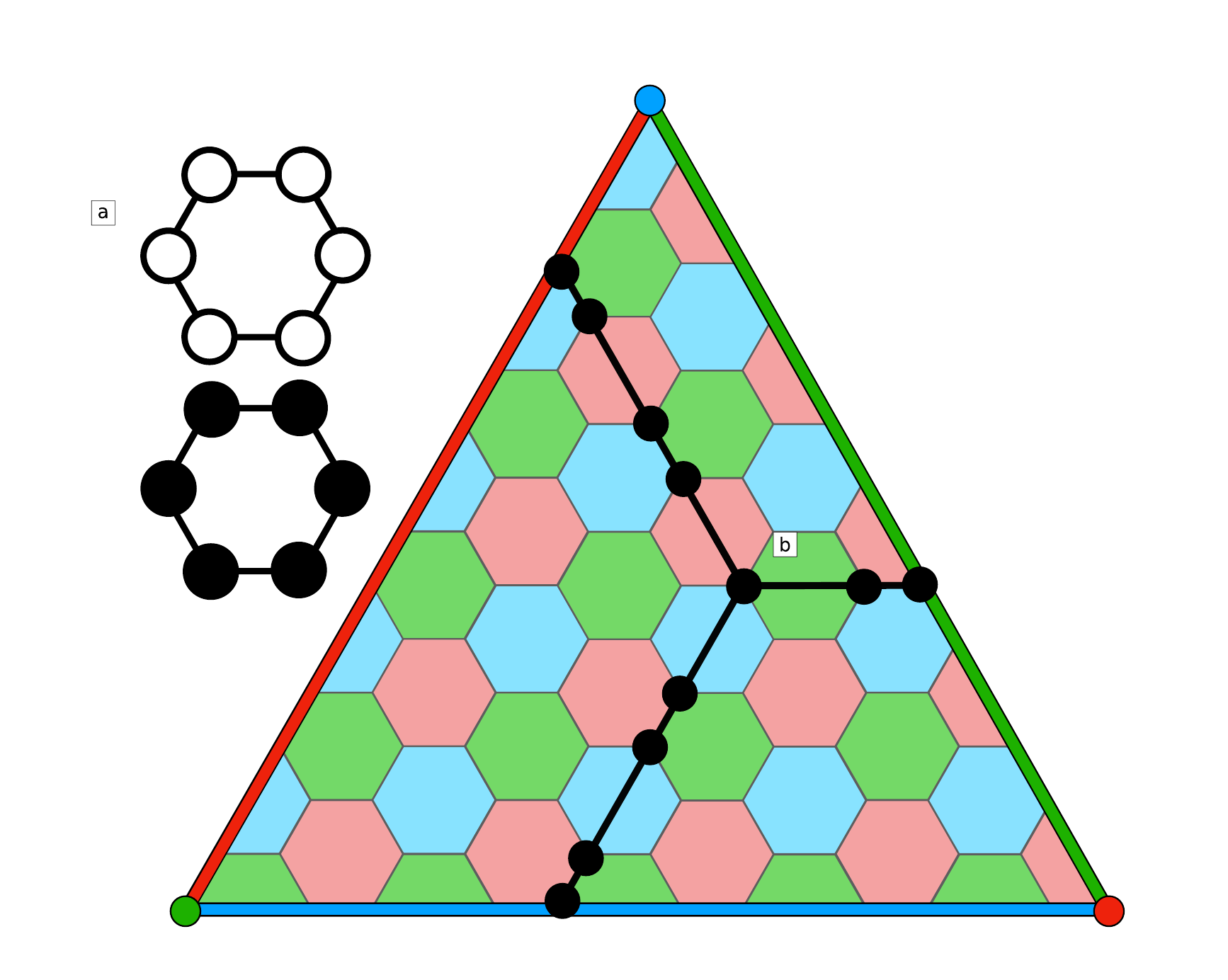}
    \caption{Triangular color code on a hexagonal lattice with boundaries. a) The two kinds of checks of the code, b) a logical operator of the code.\label{fig:tccs}}
\end{figure}

A planar quantum color code is defined on a two-dimensional lattice that is three-valent and three-colorable with respect to its faces. Let $V$ and $F$ denote the set of vertices and faces of the lattice, respectively. The qubits are placed on the vertices $v \in V$ and checks are associated with each face of the lattice $f\in F$. Each check acts on the qubits around it, essentially computing the parity of the qubits. Besides the checks, a crucial set of operations on the code are \emph{logical operators} (which are analogous to codewords in classical coding theory). It is important to note that logical operators change the encoded information, which is vital when the decoding problem is considered. For triangular color codes, logical operators that act on the encoded information pictorially look like strings that (i) run along a side of the triangle, (ii) connect a boundary to the opposite vertex of the same color, or (iii) strings that connect all three boundaries of the triangle.
\begin{example}
    An important family of color codes is the so-called 6.6.6 triangular color code, which is defined on a hexagonal lattice with a triangular shape (with boundaries), as depicted in~\Cref{fig:tccs} (for $d=11$).
    Checks involving four or six qubits are associated with the faces of the lattice and the qubits are put on the vertices of the lattice. The checks associated with the hexagonal faces act on the qubits around the faces. A logical operator acting on 11 qubits is shown. 
\end{example}
In general, the distance $d$ of a quantum CSS code is the minimum Hamming weight of a non-trivial logical operator. In the case of triangular color codes, for example, the distance is given as the number of qubits in a string along a side of the triangle. The overall parameters of the triangular color code are $[[(3d^2+1)/4, 1, d]]$. Hence, the distance can be increased by simply considering triangles featuring larger side lengths.
In case of an error occurring on a qubit, the checks of the adjacent faces are used to ``detect'' the error and the collection of faces that are ``triggered'' by an error is called the \emph{syndrome} of an error. Pictorically this can be depicted as highlighting the triggered faces accordingly. Thus, an error on a single qubit triggers the adjacent faces around it. 

\section{Decoding color codes}\label{sec:overview}
In this section, an overview of the considered problem and a brief review of prior work on decoding color codes is given.
\vspace*{-1mm}
\subsection{Considered problem}\label{sec:considered_problem}
In the remainder of this work, we consider the decoding problem for quantum color codes and for that, we use the  triangular color codes as a suitable representative. 
However, the proposed decoder is applicable to any color code and can even be applied to CSS codes in general.
We focus on the bit/phase-flip noise model, where each qubit is independently affected by an error with probability~$p$---the \emph{physical error rate}. 
Since each face of the lattice corresponds to a check, an error on a qubit (a vertex in the lattice) leads to a violation of adjacent checks. Note that this is a rather simple noise model. However, it is well-established for determining the principle capabilities of codes and decoders.

An error on $n$ qubits can be represented as a binary vector $e \in \mathbb{F}_{2}^n$, which is simply the indicator vector for the support of the error, i.e., $e_i = 1$ iff qubit $i$ is affected by the error. The \emph{syndrome} $s = H \cdot e$ corresponds to a vector indicating the violated checks (faces on the lattice) and is used as input to the decoder. The goal of the decoding procedure is to infer an estimate $\varepsilon$ for $e$ that is consistent with the syndrome, i.e., $s = H\cdot \varepsilon$. In fact, for quantum codes it is only required to find an estimate \emph{up to stabilizer}, which means that the residual error $r = e+\varepsilon$ is in the rowspace of the parity-check matrix. Moreover, to reduce the probability of a logical error, it is required that the estimate $\varepsilon$ involves a minimal number of qubits, i.e., has minimal support.
 \begin{figure}
     \centering
     \begin{subfigure}[b]{0.38\textwidth}
         \centering
         \includegraphics[width=0.6\linewidth]{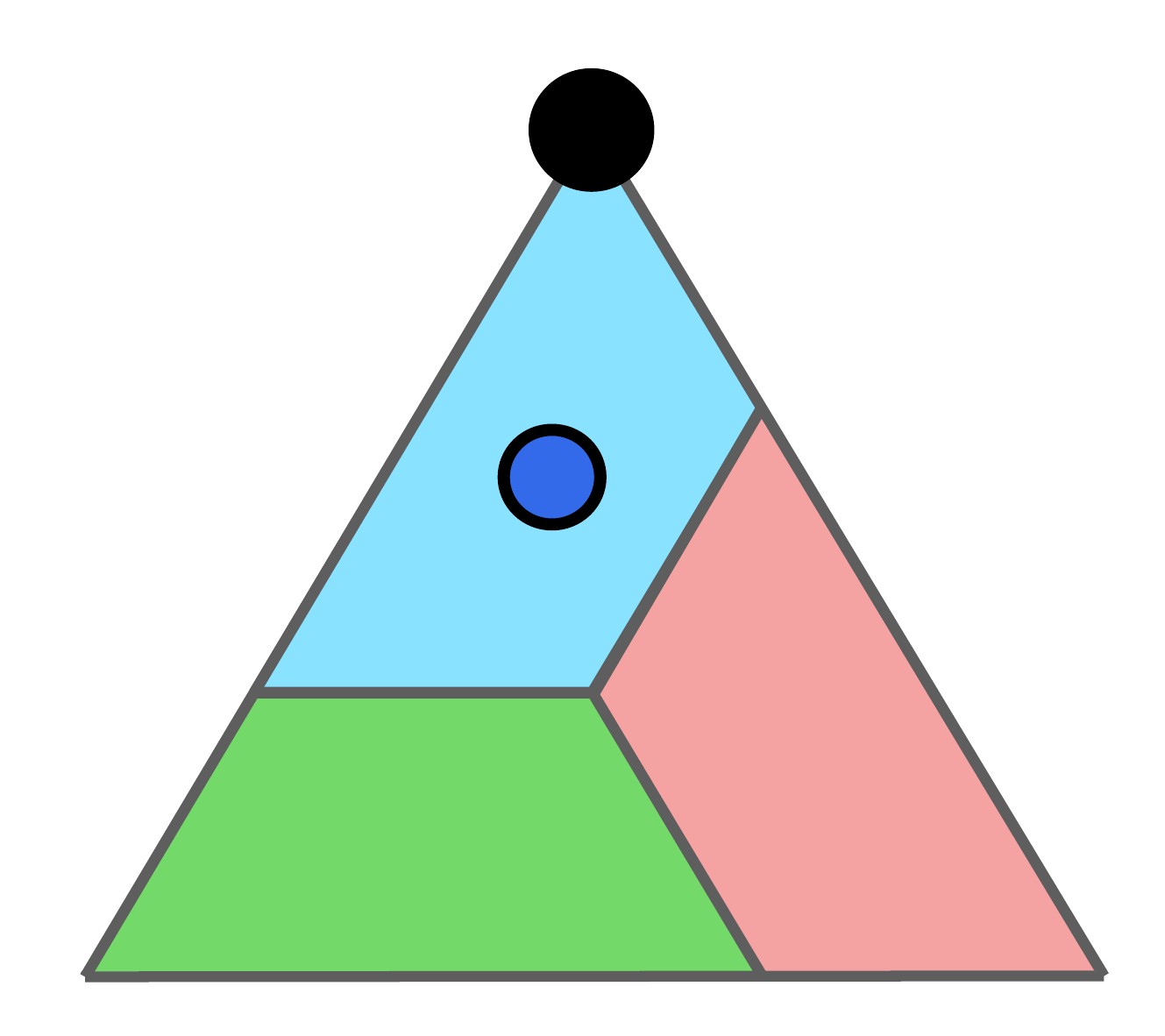}
        \label{fig:cc-dec-example-a}
     \end{subfigure}
    \hfill
     \begin{subfigure}[b]{0.4\textwidth}
         \begin{align*}
            & H = \left(\begin{matrix}
				1 &0 &0 &1 &0 &1 &1 \\
				0 &1 &0 &1 &1 &0 &1 \\
				0 &0 &1 &0 &1 &1 &1\\
			    \end{matrix}\right),\\
             &e = (1,0,0,0,0,0),\\
             &s = H\cdot e = (1,0,0),\\
             &\text{Goal: find } \varepsilon \text{ such that } H \varepsilon = s.
         \end{align*}
        \label{fig:cc-dec-example-b}
     \end{subfigure}\vspace*{-8mm}
    \caption{Example of a single-qubit error on a distance-$3$ triangular color code.\label{fig:steane-ex}}
\end{figure}
\begin{example}\label{ex:steane}
    Consider again the parity-check matrix from~\Cref{ex:hammingc}. This corresponds to a parity-check matrix of a $7$-qubit triangular color code, as depicted in~\Cref{fig:steane-ex}. Assume that the qubit ordering induces index $0$ for the qubit on top of the triangle and the indices of the faces are $\mathit{blue}=0$, $\mathit{green}=1$, and $\mathit{red}=2$. A single-qubit error happens on the top-most qubit of the triangle, which leads to a violation of the check on the adjacent face, (indicated by the blue circle). The error on the first qubit corresponds to the binary vector shown on the right-hand side of the figure. The syndrome vector then indicates that the blue face is ``triggered'' by the error. The result of all other checks is equal to $0$.
\end{example}\vspace*{-5mm}

\subsection{Related work}\label{sec:rel-work}

Several decoders for (triangular) color codes exist to date. Apart from generic decoders for CSS codes such as localized statistics decoding~\cite{hillmann2024localized}, union-find~\cite{delfosse2021almost,delfosse2022toward}, and the trellis decoder~\cite{sabo2024trellis},
one of the most studied algorithms (including several extensions and variations) is the projection decoder~\cite{delfosse2014decoding, kubica2018abcs, maskara2019advantages,bombin2012strong,beverland2021cost,li2018fault, stephens2014efficient}. 
The main idea of the projection decoder is to map the color code onto copies of surface codes 
\cite{Unfolding}
and to decode these separately using a matching-based decoder, which pairs up syndromes using the lowest-weight correction possible. Similarly, in Ref.~\cite{sahay2022decoder}, a 
\emph{minimum-weight perfect matching} approach has been applied to lattices embedded on a M\"obius strip, which has recently been implemented in~\cite{gidney2023new} while this manuscript was in preparation for publication. 
Similarly, in Ref.~\cite{lee2024color}, a matching-based approach was proposed that combines two matching decoders per color. 
Other types of decoders, whose performance for the color code has been benchmarked, are \emph{neural network decoders}~\cite{maskara2019advantages} and \emph{renormalization group decoders}~\cite{sarvepalli2012efficient}.
A slightly different approach, which has been introduced in Ref.~\cite{miguel2022cellular}, makes use of a cellular automaton decoder \cite{CADecoders} for a variant of the color code for biased noise. 

In comparison to existing decoders, an important aspect of the proposed MaxSAT decoder is that it is versatile and can be applied to an arbitrary color code (in fact, an arbitrary quantum CSS code).
A clear limitation of using an underlying MaxSAT solver to do the desired computation is that there are, in general, no clear runtime guarantees.
In the worst-case, the runtime is exponential (under standard hardness assumptions in computational complexity).
A potential solution is to set a timeout for the decoder, as is done in, for instance, belief-propagation decoding~\cite{gallager1962low, kschischang2001factor}.
The timeout could be, depending on the underlying MaxSAT solving algorithm, a maximum number of iterations or similar.
On the other hand, as we observe in the numerical simulations presented below, the runtime of the proposed MaxSAT decoder scales with the physical error rate.
As the low error rate regime is where the decoder would be applied in practice, this is a highly favorable property.

\section{LightsOut analogy}\label{sec:proposed-method}

In the following, we demonstrate how the problem of decoding color codes (under the considered noise model) can be reduced to a variation of the well-known \LO{} puzzle~\cite{gervacio2011note,anderson1998turning,amin1998parity,araujo2000turn,fleischer2013survey} on the respective color code lattice.
An instance of \LO{} consists of a lattice whose faces are associated with \emph{switches} and \emph{lights} that can either be \emph{on} or \emph{off}. 
Toggling a switch on the lattice (considered a single \emph{move}) toggles all neighbouring lights, i.e., any adjacent light that was on before toggling the switch is off afterward, and vice versa.
The goal of the puzzle is, given an initial configuration of lights, to find a sequence of moves---called a \emph{solution set}---that turns off all the lights.

\begin{figure}[th]
    \centering
    \includegraphics[width=0.5\textwidth]{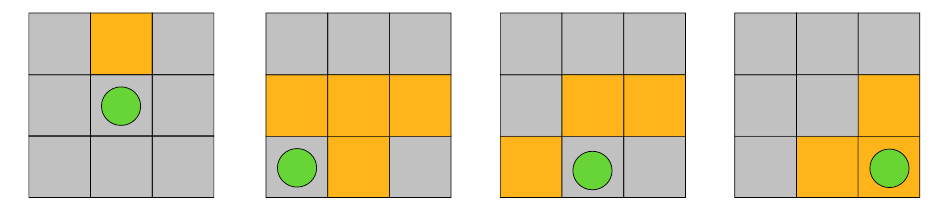}
    \caption{\LO{} puzzle and a solution on a $3\times3$ square lattice.}
    \label{fig:lo}
\end{figure}

\begin{example}
Consider the \LO{} puzzle on a $3\times3$ square lattice as illustrated in~\Cref{fig:lo}. 
In the initial configuration (the left-most lattice), a single light indicated as a yellow box, is turned on. 
Toggling the switch of the middle light, indicated by a green circle, turns on the middle light and all adjacent lights, while the light that was previously on is turned off.
Moving on, as illustrated in the figure, eventually turns off all the lights and, thus, yields a solution to the \LO{} puzzle with four moves.
\end{example}

This type of puzzle has two interesting properties~\cite{anderson1998turning}:

\begin{enumerate}
    \item Toggling a switch twice is the same as not touching it at all. As a result, each switch has to be toggled at most once for any solution to a given \LO{} puzzle.
    \item The state of a light only depends on how often the corresponding switch and its neighbors have been toggled, i.e., the order in which the switches are toggled does not matter.
\end{enumerate}

Based on these observations, it can be concluded that any solution to a \LO{} puzzle must toggle the lights that are \emph{on} in the initial configuration an \emph{odd} number of times and the lights that are initially \emph{off} an \emph{even} number of times.
The \LO{} puzzle on an $n\times m$
grid can be solved with Gaussian elimination in
polynomial time, in fact, 
in $\mathcal{O}(n^2)$ time and $\mathcal{O}(nm)$ space, where $n\leq m$ \cite{anderson1998turning}. Using this insight, one can also bound the runtime of the decoder in a polynomial fashion. For general graphs finding a minimum-weight solution to the \LO{} puzzle is NP-hard in worst-case complexity  \cite{berman2021lights}.

Now consider an instance of the decoding problem for a color code $\mathcal{C}$ given the syndrome $s = He$ of an (unknown) error $e$. Then, the question of computing an estimate $\varepsilon$ of $e$ such that $H \varepsilon = s$ can be turned into a variation of the \LO{} puzzle described above as follows:
Instead of identifying lights and switches with the same entities (i.e., the faces of the underlying lattice) consider the following variation: 

\begin{itemize}
    \item Each face of the color code lattice corresponds to a light and
    \item each vertex (qubit) of the lattice corresponds to a switch that toggles all adjacent lights.
\end{itemize}

\begin{figure}
    \centering
    \includegraphics[width=0.5\textwidth]{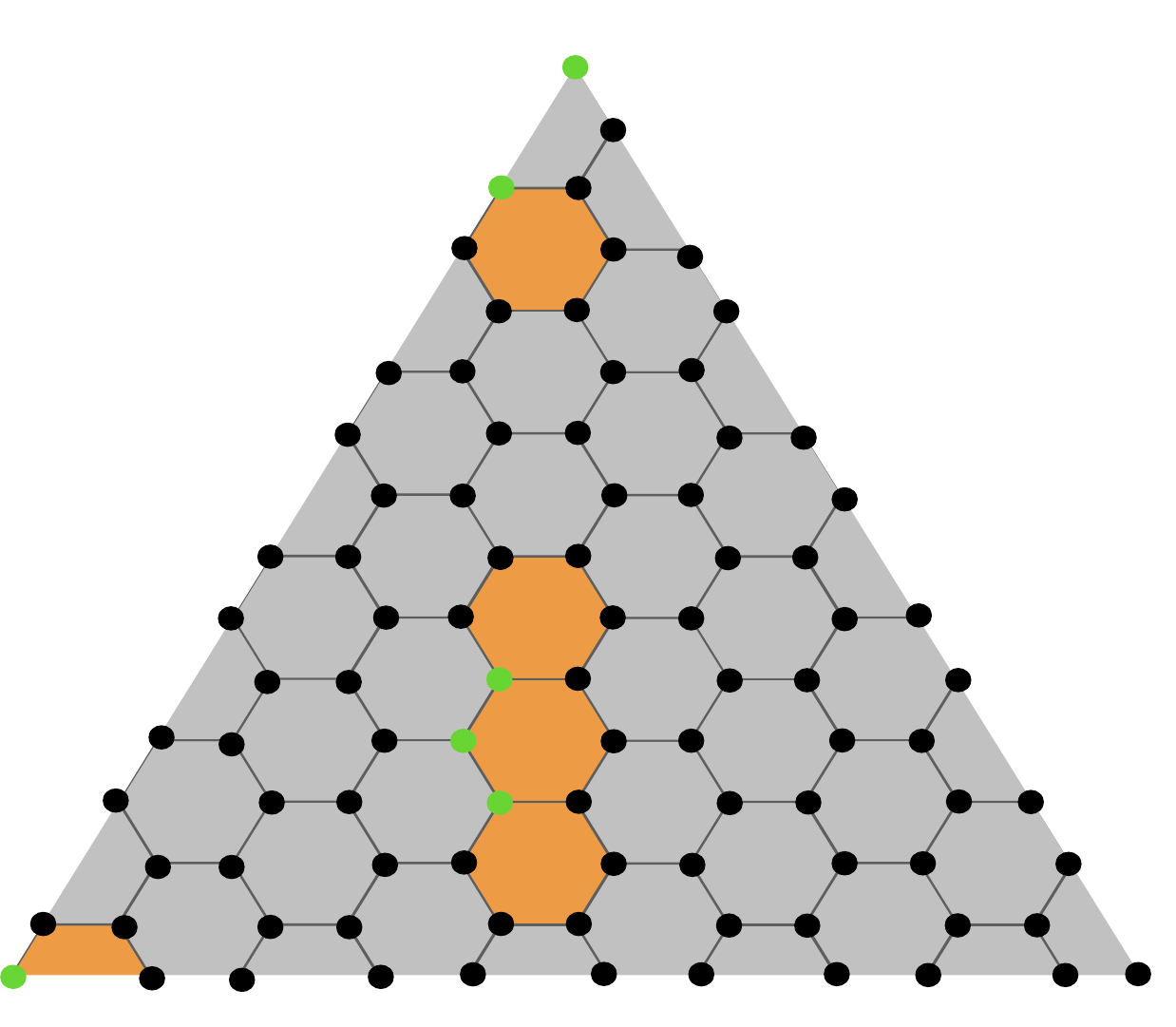}
    \caption{Color code decoding as an instance of the \LO{} puzzle. The lattice for the $d=11$ triangular color code is shown. An initial configuration is indicated by marked faces. A possible (not necessarily minimal) solution set is marked in green.}
    \label{fig:problem-form}
\end{figure}

\noindent Moreover, choose and fix an ordering of the lights and switches such that the set of lights that are turned on can be described as a binary vector whose length is equal to the number of lights (and analogously for switches).
The syndrome $s$ describes the initial configuration of the lights in the puzzle.
\begin{example}
    Consider the \LO{} illustration for the $d=11$ color code lattice depicted in~\Cref{fig:problem-form}. 
    Vertices correspond to switches, while faces correspond to lights.
    The figure shows an initial configuration with five lights on (highlighted in yellow).
\end{example}

Although the problem formulation is slightly different from the classical \LO{} puzzle, the observation made above still holds.
As such, the goal for the decoding problem in this analogy is to find a set of switches $\varepsilon$ such that toggling them turns off all the lights.
Such a set corresponds to an estimate $\varepsilon$ that is consistent with the syndrome, i.e., $H \varepsilon = s$.
Here, the size of the set corresponds to the \emph{weight} of the estimate (i.e., the number of qubits involved).
As described in \Cref{sec:considered_problem}, the goal of the decoding process is to determine an estimate with minimal (Hamming) weight. The following table provides a summary of the \LO{} analogy described throughout this section:

\begin{center}
    \begin{tabular}{l  l}%
        \toprule
             QEC & \LO{}\\
         \midrule
                Qubits/vertices             & Switches   \\
                Checks/faces             & Lights \\
             Syndrome           & Initial configuration \\
               Valid estimate     & Solution set  \\
               Min-weight estimate  & Minimal solution set \\
        \bottomrule
    \end{tabular}\\\vspace{3mm}
\end{center}
\section{MaxSAT decoding}\label{sec:maxsat} %

In the following, we propose a MaxSAT formulation of the \LO{} analogy for the color code decoding problem that allows to determine \emph{minimum-weight} solutions.
Note that even though the \LO{} description above assumes a bit/phase-flip noise model with perfect syndrome measurements, the analogy can readily be applied to more general noise models, where the \LO{} puzzle corresponds to a three-dimensional stack of the planar version with additional switches between two consecutive layers that represent so-called ``time-like'' errors.
For the sake of simplicity, we focus on the simpler, bit/phase-flip noise model in the following and discuss the more general case in~\Cref{sec:app_3d-lo}.

\subsection{General idea}\label{sec:general_idea}

In order to formulate the \LO{} analogy of the decoding problem as a MaxSAT problem, a description of the underlying lattice $(F, V)$ is needed.
To this end, we first introduce Boolean variables $\mathit{switch}_1, \dots, \mathit{switch}_{|V|}$, where $|V|$ denotes the number of vertices in the lattice, associated with the number of qubits in the code.
In addition, a discrete function $\mathcal{F}_{\mathit{switches}}\colon\;F\to V^*$ is introduced (and realized as a dictionary), that takes a face $f\in F$ as input and returns the set of vertices $\mathcal{F}_{\mathit{switches}}(f)=\{v_1,\dots,v_{n_f}\}$ surrounding the face.
Finally, a Boolean function $\mathcal{F}_{\mathit{init}}\colon\;F\to \{\mathit{true},\,\mathit{false}\}$ describes the syndrome that has been measured and shall be decoded.
In the \LO{} analogy, $\mathcal{F}_{\mathit{switches}}$ returns the switches that toggle a given light, while $\mathcal{F}_{\mathit{init}}$ describes the initial configuration of the lights.

As observed above, any valid solution to the \LO{} puzzle toggles an \emph{odd} number of switches around a light that is \emph{on} in the initial configuration, while an \emph{even} number of switches surrounding a light that is initially \emph{off} is toggled.
This can be formulated as \emph{parity constraints}

\begin{equation}
    \forall{f\in F}\colon\; \bigoplus_{v\in \mathcal{F}_{\mathit{switches}}(f)} \mathit{switch}_v = \mathcal{F}_\mathit{init}(f),
\end{equation}
where $\bigoplus$ denotes the \emph{exclusive-or}  (XOR).
Satisfying these constraints, i.e., solving the satisfiability problem under these constraints, guarantees a valid solution to the \LO{} puzzle and, hence, the decoding problem.
Moreover, the formulation can be easily adapted to a \emph{maximum satisfiability problem} (MaxSAT) by adding the following soft constraints
\begin{equation}
    \forall v\in V\colon\;\mathit{not}(\mathit{switch}_v).
\end{equation}
The general MaxSAT problem is {\tt NP}-hard
in worst-case complexity, as its solution leads to the solution of the Boolean satisfiability problem, which in turn is 
{\tt NP}-complete. 
Maximizing over these soft constraints \emph{minimizes} the switch variables that are set to \emph{true} in the solution and, hence, yields a \emph{minimum-weight} estimate for the decoding problem.

\subsection{Implementation details}\label{sec:details}

The encoding in the previous section has a rather simple form. For a given lattice $(V, F)$ representing some code, it introduces $|V|$ Boolean variables and $|F|$ parity constraints that involve at most
\begin{equation}
    \max_{f\in F} |\mathcal{F}_{\mathit{switches}}(f)|
\end{equation}
variables.
For the example of the triangular color code discussed before in \Cref{sec:background}, each parity constraint involves either four or six variables (due to the hexagonal tiling of the triangle).
However, it is well known that the way in which multi-variable parity constraints are realized heavily influences the performance of SAT solvers~\cite{soos2009extending, haanpaa2006hard, laitinen2012extending}.

Thus, for the resulting decoder to be efficient, a suitable encoding must be found.
Furthermore, the decoding problem is not a one-and-done task, but rather has to be solved over and over with different syndromes as input during the operation of an error-corrected quantum computer.
Hence, it is also important to allow for efficiently re-using an existing SAT instance instead of always constructing a new instance from scratch.
The straight-forward linear encoding is given by

\begin{widetext}
    \begin{equation}
         \left[\bigoplus_{v\in \mathcal{F}_{\mathit{switches}}(f)} \mathit{switch}_v = \mathcal{F}_\mathit{init}(f)\right] \equiv \left[\left(\left(\left(\mathit{switch}_1 \oplus \mathit{switch}_2\right) \oplus \mathit{switch}_3\right)\cdots\right) = \mathcal{F}_\mathit{init}(f)\right]
    \end{equation}
\end{widetext}

satisfies none of the above criteria. Such an encoding is known to cause problems for SAT solvers and there is no direct way of adjusting these constraints to new values of $\mathcal{F}_\mathit{init}(f)$ despite recreating them as a whole.

To overcome this issue, the proposed formulation introduces $k=|\mathcal{F}_{\mathit{switches}}(f)|-1$ helper variables $h_1,\dots,h_{k}$ and splits the overall constraint into $k+1$ separate constraints
\begin{align}
    \mathit{switch}_1 \oplus h_1 &= \mathcal{F}_\mathit{init}(f), \nonumber \\
    h_{i} &= \mathit{switch}_{i+1} \oplus h_{i+1} , \\
    h_{k} &= \mathit{switch}_{k}, \nonumber
\end{align}
for  $i=1,\dots, k-1$.
It is easy to verify that this still realizes the parity constraint on the switch variables.
Observe how only the first of those constraints depends on the value of $\mathcal{F}_\mathit{init}(f)$.
All other constraints can be pre-computed independently.
As confirmed by numerical evaluations, which are summarized next, this yields an efficient and re-usable MaxSAT decoder for quantum color codes that achieves state-of-the-art decoding performance (for the considered noise model and code).

\section{Numerical evaluation}\label{sec:experiments}
To evaluate the proposed approach, the decoder performance, as well as the runtime (scaling) of the proposed approach, have been comprehensively investigated using numerical simulations. 
The results will be presented in two parts. First, the decoder performance and the \emph{threshold}---a well-established figure of merit---are discussed. 
In a second series of experiments, we focus on the runtime scaling of the proposed approach.
In addition to all numerical data, the implementation of the decoder as well as the means to conduct numerical simulations are made available at
\url{www.github.com/cda-tum/qecc}.

\begin{figure*}[tbh]
    \centering
    \includegraphics[width=0.99\textwidth]{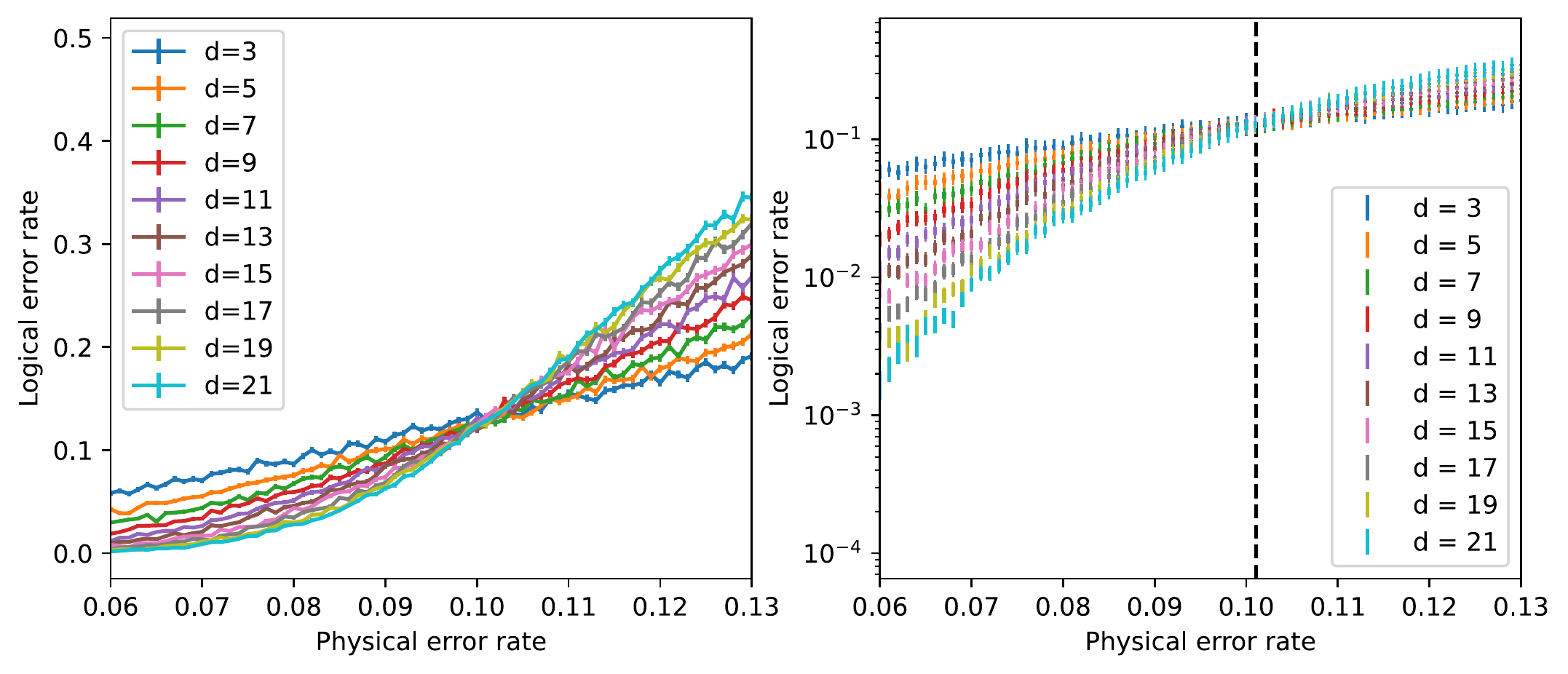}
    \caption{Logical error rates of triangular color codes with distances ranging from 3 to 21 under bit-flip noise. The right plot shows the logical error rate scaling on a logarithmic y-axis around the threshold of $p_{th}\approx 10.1\%$. The error bars in the right plot correspond to one standard deviation. \vspace*{-5mm}}
    \label{fig:lers-th}
\end{figure*}

\subsection{Decoding performance}\label{sec:dp}

In this section, findings related to the decoding performance are presented. 
To this end, the proposed decoder has been implemented using the MaxSAT engine of Microsoft's Z3~\cite{de2008z3} solver. All simulations have been conducted on a machine equipped with an \mbox{AMD Ryzen 9 5950X} CPU and \SI{64}{\gibi\byte} RAM running Ubuntu~22.04. 
As previously stated in the paper, the error model under consideration is \emph{bit-flip noise}.
In a first round of experiments, the \emph{logical error rate} of triangular color codes is investigated. Recall that a logical error is an error that changes the logical information. Thus, if a decoding estimate (the resulting residual error) induces a logical error, the decoding process is considered to have failed, since the proposed estimate (together with the actual error) has altered the logical information. 
A widely used figure of merit of a code $+$ decoder $+$ error model combination is the \emph{threshold} $p_{th}$. Intuitively, this is an estimate of a physical error rate up to which the code is beneficial. For $p<p_{th}$, the logical error rate can be exponentially suppressed by scaling the code (inducing a potentially large overhead). The optimal threshold of color codes on a hexagonal lattice under bit-flip noise is $\approx 10.9\%$~\cite{katzgraber2009error}, which is achieved by the tensor network decoder~\cite{chubb2021general} as mentioned in~\Cref{sec:rel-work}.

In order to investigate the logical error rates and estimate the threshold, we run \mbox{Monte-Carlo} simulations for increasing physical error rates $p$ and code distances up to $d=21$. A single sample is obtained by means of the following steps: 
\begin{itemize}
    \item Sample an error $e \in \mathbb{F}_2^n$,
    \item compute the syndrome $s = H\cdot e$,
    \item use the proposed decoder to get an estimate $\varepsilon$,
    \item compute the residual error $r = e + \varepsilon$, and
    \item check if $r$ is a logical operator.
\end{itemize}
If $r$ is a logical operator, the correction, together with the original error, has altered the encoded information, which is recorded as a \emph{logical error}. The \emph{logical error rate} (LER) is then computed as the number of logical errors over the number of samples. 

The results are shown on the left-hand side in~\Cref{fig:lers-th}. 
Our results show that the proposed decoder achieves a threshold of $p_{th}\approx 10.1\%$, outperforming many other state-of-the-art decoders. The threshold is computed using the critical exponent method described in Ref.~\cite{WANG200331}. The logical error rate close to the threshold is shown on the right-hand side in \Cref{fig:lers-th}. 

In order to put this in context, the following table shows a comparison with other state-of-the-art decoders and their threshold for color codes on a hexagonal lattice under the same noise model as considered in this work.
\begin{center}
     \begin{tabular}{l r r}%
        \toprule
             Decoder & Threshold & Source\\
         \midrule
                         Optimal           & $10.9\%$  & \cite{katzgraber2009error}  \\
                         Tensor network    & $10.9\%$  & \cite{chubb2021general}\\
                         MaxSAT*           & $10.1\%$  & This work \\
                         Trellis           & $10.1$\%  & \cite{sabo2024trellis}\\ 
                         Neural network    & $10.0\%$  & \cite{maskara2019advantages} \\
                         M\"obius MWPM     & $9.0\%$   & \cite{sahay2022decoder}  \\
                         Restriction MWPM  & $8.7\%$   & \cite{delfosse2014decoding}  \\
                         Union-find        & $8.4\%$   & \cite{delfosse2021almost}  \\
                         RG                & $7.8\%$   & \cite{sarvepalli2012efficient} \\    
        \bottomrule
    \end{tabular}\\\vspace{1mm}
    \label{table:sota}
\end{center}

\begin{figure}[t]
    \centering
    \includegraphics[width=0.5\textwidth]{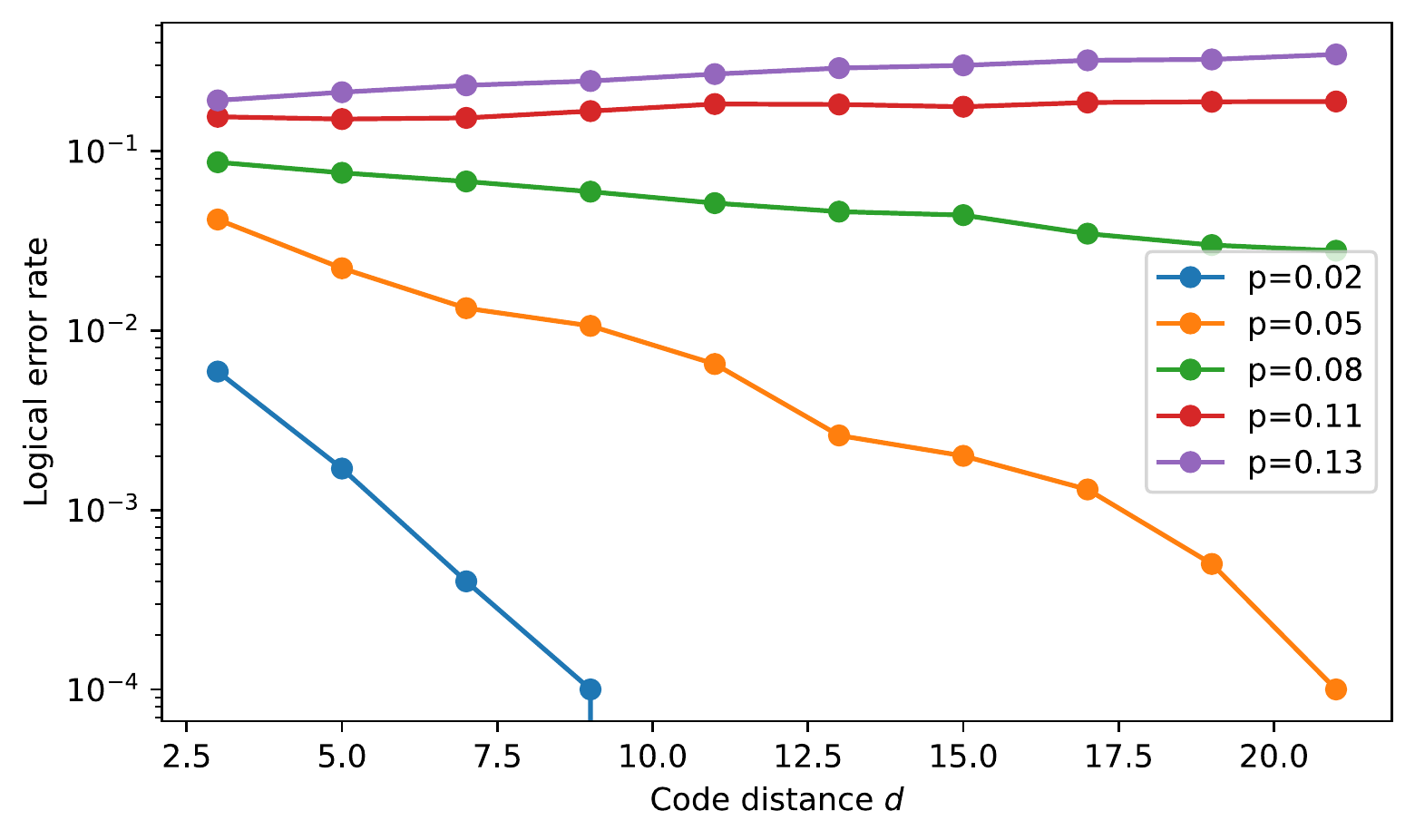}
    \caption{Logical error rate per distance for various physical error rates $p$. Note that $p=0.11$ and $p=0.13$ are above threshold.}
    \label{fig:lers-perd}
\end{figure}

A fundamental theorem of FT, the \emph{threshold theorem} intuitively states that if errors can be exponentially suppressed below some constant threshold $p_{th}$, arbitrarily long quantum computations can be conducted fault-tolerantly~\cite{aharonov1997fault, knill1998resilient, kitaev1997quantum, preskill1998reliable}. 
To demonstrate exponential sub-threshold scaling, we plot the logical error rates as a function of the distance for different physical error rates in~\Cref{fig:lers-perd}. 
For error rates below the threshold $p<p_{th}$, it can be verified that for increased distances, the logical error rate decreases exponentially, while naturally, above the threshold, this is not the case.

Recall that the distance of an $[[n,k,d]]$-code is the minimum weight of a (non-trivial) logical operator. In general, the distance is an indicator of how many errors $t$ a code can, in principle, correct: $t\leq (d-1)/2$. In the best case, the decoder performance matches this upperbound. In our numerical experiments, we observe that the minimum weight logical operators induced by the decoder always have weight equal to the distance $d$ for the investigated distances. Note that it is common for decoders to be only able to correct up to a polynomial fraction of $d$. 

\subsection{Runtime scaling}\label{sec:rt}
\begin{figure*}[t]
    \centering
    \includegraphics[width=0.89\textwidth]{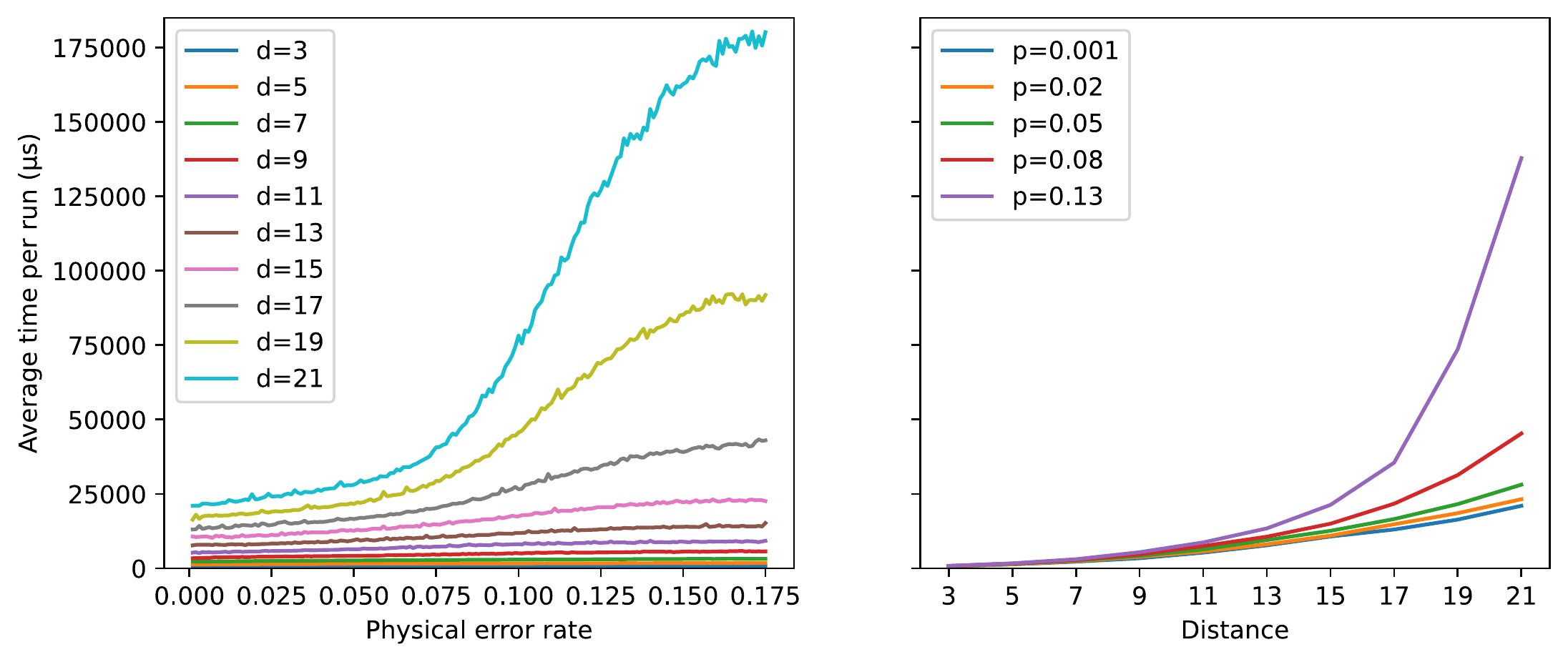}
    \caption{Average runtime in microseconds for a single decoding run of the MaxSAT decoder on triangular color codes and runtime per distance for several physical error rates $p$. Every data point has been obtained from $10^4$ samples.}
    \label{fig:rts}
\end{figure*}

To estimate the time needed to decode and to determine how the runtime performance scales as the codes under consideration get larger, numerical simulations to measure the time needed to solve the MaxSAT instance were conducted. 
Since the MaxSAT instance and most of the needed information concerning the code can be precomputed, the dominant bottleneck of a single correction cycle is the MaxSAT solver. Thus, we only record the time needed to solve a given instance, which we indicate as \emph{runtime} of the solver.
The results are summarized in~\Cref{fig:rts}. Overall the runtime for a single decoding takes up to a hundred milliseconds for distances smaller than $21$ and a few hundred milliseconds for larger distances. What is especially interesting is that the runtime scales proportionally with the physical error rate, hence for lower physical error rates the runtime is better. This is similar to the runtime behaviour of matching-based decoders and beneficial since in practice the lower physical error rates are the more relevant ones.

To put the obtained estimates in context (without arguing that the proposed approach is directly applicable to be employed for physical devices), consider that for realizations of quantum error-correction, a quantum device built on a superconducting architecture has cycles times in the micro-second regime~\cite{google2021exponential,google2023suppressing}. This means that the decoder would need to be able to decode single instances in such time regimes as well in order not to slow down the whole computation. For ion-trap based devices, it is known that operations are slower and cycle times are in the millisecond regime~\cite{ryan2021realization}. Note that for such practical settings, the decoder needs to be able to take a more involved noise model into account. Additionally, since the runtime is highly dependent on the MaxSAT formulation, how the constraints are exactly formulated, and the solver itself, we expect that the obtained runtimes can be optimized further. However, we leave more fine-grained runtime optimizations for future work.

\begin{figure}[tbh]
    \centering
    \includegraphics[width=.45\textwidth]{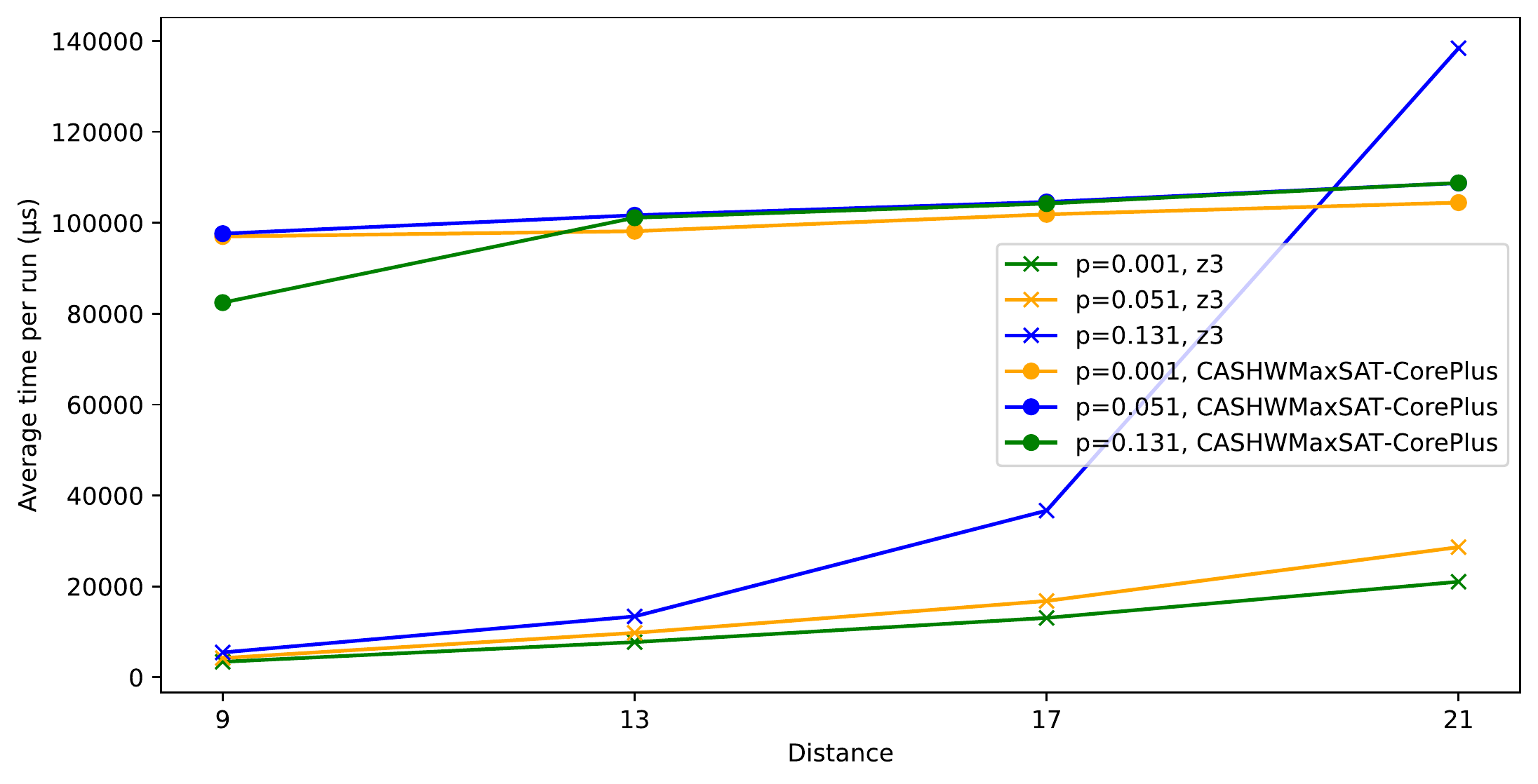}
    \caption{Comparison of the runtime performance of a state-of-the-art MaxSAT solver for different physical error rates on triangular color codes.}
    \label{fig:rt-cmp-solvers}
\end{figure}

Since MaxSAT solvers and their optimization are vibrant areas of research, we conducted simulations to compare the original implementation that uses Z3 and other state-of-the-art MaxSAT solvers. 
More specifically, we conducted simulations to compare the runtime performance of all MaxSAT solvers that participated in the \emph{MaxSAT Evaluation 2022}  (MSE22)~\cite{bacchus2022maxsat}.
Preliminary evaluations indicated that all solvers from the MSE22 have comparative runtime performance---running within $\approx 4\%$ of each other.
Consequently, we chose the winner of the MSE22 (\emph{CASHWMaxSAT-CorePlus}) and compared its runtime against the original implementation that uses Z3.
The results, which are depicted in \Cref{fig:rt-cmp-solvers}, show that the Z3-based solution outperforms the other MaxSAT solver in all but one case.
The most vital insight, however, is that only the performance of the original implementation that uses Z3 as solver has runtime performance that scales proportionally with the physical error rate. 
This is not true for all other investigated solvers. 
Since the sub-threshold regime (low physical error rate) is the one relevant in practice, this behaviour is desirable and, overall, the Z3-based implementation outperforms all other solvers in this regime.
One reason for this phenomenon could be that Z3 (as an SMT solver) performs a diverse set of optimizations on the constructed MaxSAT formulations, while, in order to use the other MaxSAT solvers, the formulation had to be converted to \emph{weighted CNF} (WCNF)---for which we used the automatic conversion methods built into Z3).
This conversion might limit the potential for optimization that is performed by the solvers.

\subsection{Comparison with the tensor network decoder}

To draw a comparison to implementations of existing decoders, we compare the runtime and the decoding performance of the proposed MaxSAT decoder to the \emph{tensor network decoder}~\cite{chubb2021general}, as this is the only known state-of-the art decoder that has a higher threshold than the proposed MaxSAT decoder. We use the tensor network decoder implementation provided in the QECSIM tool~\cite{qecsim} (which---to the best of our knowledge---is the only open-source implementation).

\begin{figure*}[t]
    \centering
    \includegraphics[width=.9\textwidth]{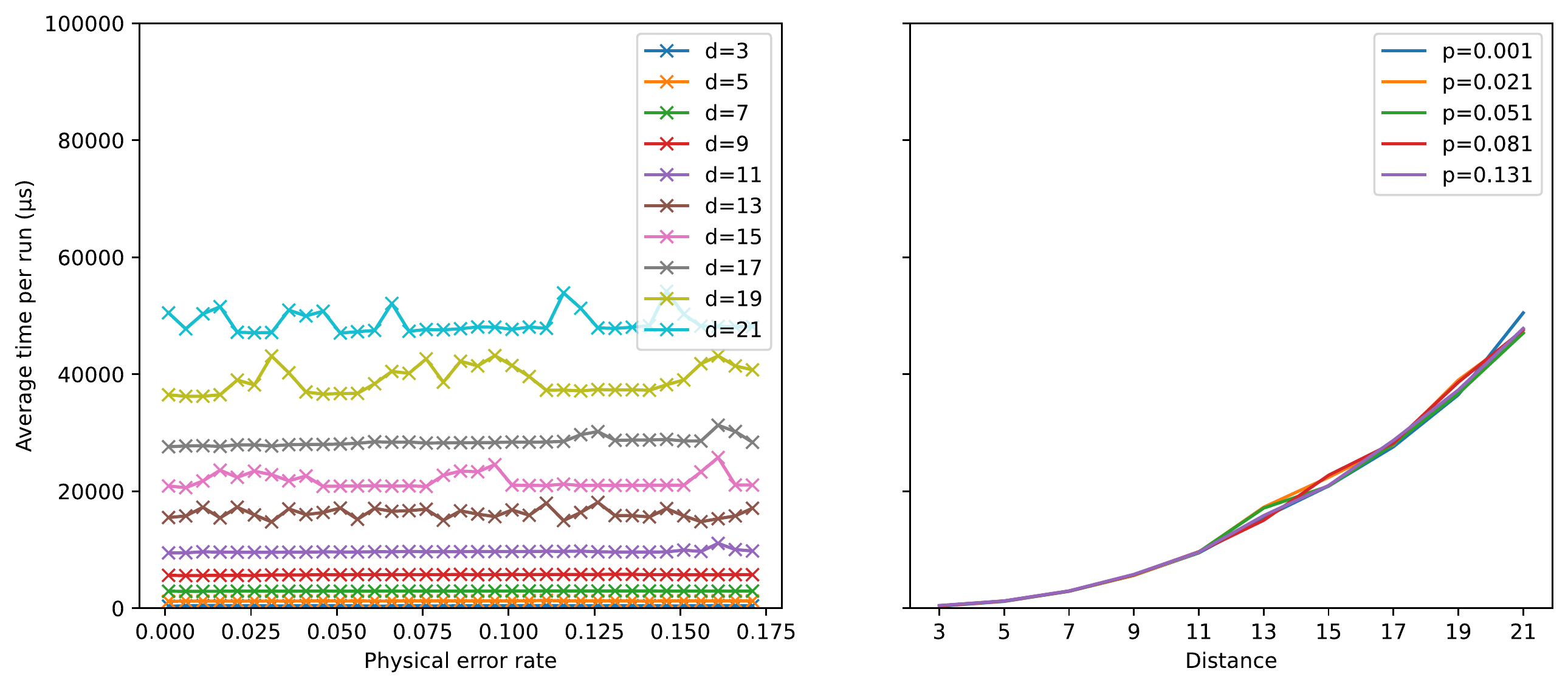}
    \caption{Runtime performance of the tensor network decoder. The results are obtained using $10^4$ samples and the maximum bond dimension of the tensor network is set to 6.}
    \label{fig:rts-tn}
\end{figure*}

We run the decoder on the same machine used for the MaxSAT decoder simulations. The decoder takes as an input parameter the maximum bond dimension, which we set to $\chi=6$.
Note that, however, both the logical error rate and the runtime of the TN decoding algorithm depend on $\chi$. 
We have investigated the logical error rate saturation of the TN decoder and have chosen $\chi$ accordingly. A more in-depth discussion in presented in~\Cref{sec:app_tn-cmp}. 
The runtime includes the time QECSIM takes to sample an error and check if a logical error occurred, but the obtained estimates are enough to draw reasonable comparisons, as the overall time is dominated by the decoding time.
The results are shown in~\Cref{fig:rts-tn}. 
Our results indicate that the runtime of the tensor network decoder does not scale with the physical error rate. This agrees with the known runtime of the tensor network decoder, which is $\mathcal{O}(n \chi^3)$, where $n$ is the number of qubits and $\chi$ is the bond dimension.  
The MaxSAT decoder runtime does scale with physical error rate and it therefore performs better for sub-threshold error rates. 
To be specific, for a distance $21$-code at a physical error rate of $10^{-3}$ the MaxSAT decoder is approximately $2$ times faster than the tensor network decoder ($\approx 25$ vs $\approx 50$ milliseconds).

We are unable to compare the runtime of the MaxSAT decoder to that of the trellis decoder, as there is no open-source implementation of the latter. 
Similar to the tensor network decoder, the runtime of the trellis decoder does not scale with the physical error rate~\cite{sabo2024trellis}. Therefore, we expect that although the threshold of the MaxSAT decoder is equivalent to that of the trellis decoder, it is favorable to use the former due to its expected shorter runtime for low physical error rates.
We leave a numerical investigation comparing the runtimes of the two decoders to future work.
Furthermore, we emphasize that the runtime comparisons presented here are not fully representative and strongly depend on the 
implementation and underlying hardware.
Hence, the presented results are not meant to reflect the expected runtime of the decoder for a hardware experiment. 
The conducted numerical simulations are merely meant to provide a rough estimation of the decoder's behaviour and relate it to an existing and publicly available implementation.

\vspace*{-2mm}
\section{Conclusions and outlook}\label{sec:conclusion}
\vspace*{-2mm}
In this work, we have proposed a novel decoding approach for quantum codes. Based on an analogy of the decoding problem for color codes and the well-known mathematical puzzle, \LO{}, we have introduced a MaxSAT solution to the corresponding problem. Moreover, an efficient construction of the corresponding satisfiability instance through various constraint optimizations has been proposed. In several numerical simulations, it has been shown that the proposed decoder achieves near-optimal decoding performance. Moreover, the experimental evaluation signifies that the runtime performance of the decoder scales proportionally with the error rate and thus outperforms state-of-the-art decoders, such as the tensor network decoder for low error rates. In addition to the high performance and good runtime scaling, an advantage of the proposed decoder is that it is very general and not limited to color codes. The decoder implementation, as well as all the obtained data are integrated into the open-source software package QECC~\url{https://github.com/cda-tum/mqt-qecc} as part of the \emph{Munich Quantum Toolkit (MQT)}~\cite{10646543}.

Concerning future work building on the findings laid out here, one of the most exciting directions is to extend the considered noise model to the more realistic scenario of circuit-level noise. 
It would be interesting to see what the exact limitations of the decoder performance for such noise models are. 
Moreover, since for low-weight syndromes, the runtime performance is good, a combination with other decoders such as \emph{union-find} or \emph{belief-propagation} (belief-MaxSAT, union-MaxSAT) might result in a better overall performance for larger codes. 
Furthermore, it is reasonable to assume that the runtime performance can be further improved through optimizations regarding the MaxSAT solver. 
In particular, there exist MaxSAT solvers such as \emph{GaussMaxHS}~\cite{soos2009extending} that allow optimizations for XOR constraints. 
Overall, a general goal of this work is to highlight that fundamental problems in quantum computing are still open and to foster interest in solving these problems with known and well-established tools from the field of SAT solvers and constraint satisfaction research. 
In turn, these tools may help ongoing research in tackling the decoding problem that to date is one of the roadblocks against fault tolerant quantum computing.
\vspace*{-2mm}
\section{Acknowledgements}
\vspace*{-2mm}
The authors thank Craig Gidney for fruitful discussions on the decoder and for pointing out the possible \LO{} analogy. 
Moreover, we thank Kenneth R. Brown for pointing out Ref.~\cite{sabo2024trellis} and Mate Soos for pointing out the GaussMaxHS solver reference~\cite{soos2009extending}.
L.~B., L.~B., and R.~W.\ acknowledge funding from the European Research Council (ERC) under the European Union’s Horizon 2020 research and innovation program (Quantum Flagship
Millenion, PasQuans2) as well as support by the BMWK on the basis of a decision by the German Bundestag through project QuaST. P.~J.~D.\ and J.~E.\ acknowledge funding from BMBF (RealistiQ, QSolid) and the DFG (CRC 183). This work has been conducted within the framework of the Munich Quantum Valley, which is supported by the Bavarian state government with funds from the Hightech Agenda Bayern Plus, for which this is a joint-node project.

\bibliographystyle{plainnat}
\bibliography{refs}

\begin{thebibliography}{66}
\providecommand{\natexlab}[1]{#1}
\providecommand{\url}[1]{\texttt{#1}}
\expandafter\ifx\csname urlstyle\endcsname\relax
  \providecommand{\doi}[1]{doi: #1}\else
  \providecommand{\doi}{doi: \begingroup \urlstyle{rm}\Url}\fi

\bibitem[Acharya et~al.(2023)]{google2023suppressing}
Rajeev Acharya et~al.
\newblock Suppressing quantum errors by scaling a surface code logical qubit.
\newblock \emph{Nature}, 614:\penalty0 676--681, 2023.
\newblock \doi{10.1038/s41586-022-05434-1}.

\bibitem[Aharonov and Ben-Or(1997)]{aharonov1997fault}
Dorit Aharonov and Michael Ben-Or.
\newblock Fault-tolerant quantum computation with constant error.
\newblock In \emph{Proceedings of the twenty-ninth annual ACM symposium on Theory of computing}, pages 176--188, 1997.
\newblock \doi{10.1145/258533.258579}.

\bibitem[Amin et~al.(1998)Amin, Clark, and Slater]{amin1998parity}
Ashok~T. Amin, Lane~H. Clark, and Peter~J. Slater.
\newblock Parity dimension for graphs.
\newblock \emph{Discrete mathematics}, 187:\penalty0 1--17, 1998.
\newblock \doi{10.1016/S0012-365X(97)00242-2}.

\bibitem[Anderson and Feil(1998)]{anderson1998turning}
Marlow Anderson and Todd Feil.
\newblock Turning lights out with linear algebra.
\newblock \emph{Mathematics Magazine}, 71:\penalty0 300--303, 1998.
\newblock \doi{10.1080/0025570X.1998.11996658}.

\bibitem[Ara{\'u}jo(2000)]{araujo2000turn}
Paulo~Ventura Ara{\'u}jo.
\newblock How to turn all lights out.
\newblock \emph{Elemente der Mathematik}, 55:\penalty0 135--141, 2000.
\newblock \doi{10.1007/s000170050079}.

\bibitem[Bacchus et~al.(2022)Bacchus, Berg, J{\"a}rvisalo, Martins, and Niskanen]{bacchus2022maxsat}
Fahiem Bacchus, Jeremias Berg, Matti J{\"a}rvisalo, Ruben Martins, and Andreas Niskanen.
\newblock Maxsat evaluation 2022: Solver and benchmark descriptions.
\newblock \emph{Department of Computer Science Series of Publications B; Report B-2020-2}, 2022.
\newblock \doi{http://hdl.handle.net/10138/318451}.

\bibitem[Berman et~al.(2021)Berman, Borer, and Hungerb{\"u}hler]{berman2021lights}
Abraham Berman, Franziska Borer, and Norbert Hungerb{\"u}hler.
\newblock Lights out on graphs.
\newblock \emph{Mathematische Semesterberichte}, 68\penalty0 (2):\penalty0 237--255, 2021.
\newblock \doi{10.3929/ethz-b-000483037}.

\bibitem[Beverland et~al.(2021)Beverland, Kubica, and Svore]{beverland2021cost}
Michael~E. Beverland, Aleksander Kubica, and Krysta~M. Svore.
\newblock Cost of universality: A comparative study of the overhead of state distillation and code switching with color codes.
\newblock \emph{PRX Quantum}, 2:\penalty0 020341, 2021.
\newblock \doi{10.1103/PRXQuantum.2.020341}.

\bibitem[Bomb{\'\i}n and Martin-Delgado(2006)]{bombin2006topological}
H{\'e}ctor Bomb{\'\i}n and Miguel~Angel Martin-Delgado.
\newblock Topological quantum distillation.
\newblock \emph{Physical Review Letters}, 97:\penalty0 180501, 2006.
\newblock \doi{10.1103/PhysRevLett.97.180501}.

\bibitem[Bombin et~al.(2012)Bombin, Andrist, Ohzeki, Katzgraber, and Martin-Delgado]{bombin2012strong}
H{\'e}ctor Bombin, Ruben~S. Andrist, Masayuki Ohzeki, Helmut~G. Katzgraber, and Miguel~A. Martin-Delgado.
\newblock Strong resilience of topological codes to depolarization.
\newblock \emph{Physical Review X}, 2:\penalty0 021004, 2012.
\newblock \doi{10.1103/PhysRevX.2.021004}.

\bibitem[Bonilla~Ataides et~al.(2021)Bonilla~Ataides, Tuckett, Bartlett, Flammia, and Brown]{bonilla2021xzzx}
J.~Pablo Bonilla~Ataides, David~K. Tuckett, Stephen~D. Bartlett, Steven~T. Flammia, and Benjamin~J. Brown.
\newblock {The XZZX surface code}.
\newblock \emph{Nature Communications}, 12\penalty0 (1):\penalty0 2172, 2021.
\newblock \doi{10.1038/s41467-021-22274-1}.

\bibitem[Calderbank and Shor(1996)]{calderbank1996good}
A.~Robert Calderbank and Peter~W. Shor.
\newblock Good quantum error-correcting codes exist.
\newblock \emph{Physical Review A}, 54:\penalty0 1098, 1996.
\newblock \doi{10.1103/PhysRevA.54.1098}.

\bibitem[Chubb(2021)]{chubb2021general}
Christopher~T. Chubb.
\newblock {General tensor network decoding of 2D Pauli codes}.
\newblock \emph{arXiv:2101.04125}, 2021.
\newblock \doi{10.48550/arXiv.2101.04125}.

\bibitem[De~Moura and Bj{\o}rner(2008)]{de2008z3}
Leonardo De~Moura and Nikolaj Bj{\o}rner.
\newblock {Z3: An efficient SMT solver}.
\newblock In \emph{Tools and Algorithms for the Construction and Analysis of Systems: 14th International Conference, TACAS 2008.}, pages 337--340. Springer, 2008.
\newblock \doi{10.1007/978-3-540-78800-3_24}.

\bibitem[Delfosse(2014)]{delfosse2014decoding}
Nicolas Delfosse.
\newblock Decoding color codes by projection onto surface codes.
\newblock \emph{Physical Review A}, 89:\penalty0 012317, 2014.
\newblock \doi{10.1103/PhysRevA.89.012317}.

\bibitem[Delfosse and Nickerson(2021)]{delfosse2021almost}
Nicolas Delfosse and Naomi~H. Nickerson.
\newblock Almost-linear time decoding algorithm for topological codes.
\newblock \emph{Quantum}, 5:\penalty0 595, 2021.
\newblock \doi{10.22331/q-2021-12-02-595}.

\bibitem[Delfosse et~al.(2022)Delfosse, Londe, and Beverland]{delfosse2022toward}
Nicolas Delfosse, Vivien Londe, and Michael~E. Beverland.
\newblock {Toward a union-find decoder for quantum LDPC codes}.
\newblock \emph{IEEE Transactions on Information Theory}, 68:\penalty0 3187--3199, 2022.
\newblock \doi{10.1109/TIT.2022.3143452}.

\bibitem[Dennis et~al.(2002)Dennis, Kitaev, Landahl, and Preskill]{dennis2002topological}
Eric Dennis, Alexei Kitaev, Andrew Landahl, and John Preskill.
\newblock Topological quantum memory.
\newblock \emph{Journal of Mathematical Physics}, 43:\penalty0 4452--4505, 2002.
\newblock \doi{10.1063/1.1499754}.

\bibitem[Derks et~al.(2024)Derks, Townsend-Teague, Burchards, and Eisert]{derks2024designing}
Peter-Jan~HS Derks, Alex Townsend-Teague, Ansgar~G Burchards, and Jens Eisert.
\newblock Designing fault-tolerant circuits using detector error models.
\newblock \emph{arXiv preprint arXiv:2407.13826}, 2024.
\newblock \doi{10.48550/arXiv.2407.13826}.

\bibitem[Fleischer and Yu(2013)]{fleischer2013survey}
Rudolf Fleischer and Jiajin Yu.
\newblock A survey of the game “lights out!”.
\newblock \emph{Space-Efficient Data Structures, Streams, and Algorithms}, pages 176--198, 2013.
\newblock \doi{10.1007/978-3-642-40273-9_13}.

\bibitem[Gallager(1962)]{gallager1962low}
Robert Gallager.
\newblock Low-density parity-check codes.
\newblock \emph{IRE Transactions on information theory}, 8:\penalty0 21--28, 1962.
\newblock \doi{10.1109/TIT.1962.1057683}.

\bibitem[Gervacio and Maehara(2011)]{gervacio2011note}
Severino~V. Gervacio and Hiroshi Maehara.
\newblock A note on lights-out-puzzle: Parity-state graphs.
\newblock \emph{Graphs and Combinatorics}, 27:\penalty0 109--119, 2011.
\newblock \doi{10.1007/s00373-010-0958-1}.

\bibitem[Gidney(2021)]{gidney2021stim}
Craig Gidney.
\newblock Stim: a fast stabilizer circuit simulator.
\newblock \emph{{Quantum}}, 5:\penalty0 497, 2021.
\newblock \doi{10.22331/q-2021-07-06-497}.

\bibitem[Gidney and Jones(2023)]{gidney2023new}
Craig Gidney and Cody Jones.
\newblock New circuits and an open source decoder for the color code.
\newblock \emph{arXiv:2312.08813}, 2023.
\newblock \doi{10.48550/arXiv.2312.08813}.
\newblock arXiv:2312.08813.

\bibitem[Haanp{\"a}{\"a} et~al.(2006)Haanp{\"a}{\"a}, J{\"a}rvisalo, Kaski, and Niemel{\"a}]{haanpaa2006hard}
Harri Haanp{\"a}{\"a}, Matti J{\"a}rvisalo, Petteri Kaski, and Ilkka Niemel{\"a}.
\newblock Hard satisfiable clause sets for benchmarking equivalence reasoning techniques.
\newblock \emph{Journal on Satisfiability, Boolean Modeling and Computation}, 2:\penalty0 27--46, 2006.
\newblock \doi{10.3233/SAT190015}.

\bibitem[Herold et~al.(2015)Herold, Campbell, Eisert, and Kastoryano]{CADecoders}
Michael Herold, Earl~T. Campbell, Jens Eisert, and Michael~J. Kastoryano.
\newblock Cellular-automaton decoders for topological quantum memories.
\newblock \emph{npj Quantum Information}, 1:\penalty0 15010, 2015.
\newblock \doi{10.1038/npjqi.2015.10}.

\bibitem[Higgott and Gidney(2023)]{higgott2023sparse}
Oscar Higgott and Craig Gidney.
\newblock Sparse blossom: correcting a million errors per core second with minimum-weight matching.
\newblock \emph{arXiv:2303.15933}, 2023.
\newblock \doi{10.48550/arXiv.2303.15933}.

\bibitem[Hillmann et~al.(2024)Hillmann, Berent, Quintavalle, Eisert, Wille, and Roffe]{hillmann2024localized}
Timo Hillmann, Lucas Berent, Armanda~O Quintavalle, Jens Eisert, Robert Wille, and Joschka Roffe.
\newblock Localized statistics decoding: A parallel decoding algorithm for quantum low-density parity-check codes.
\newblock \emph{arXiv preprint arXiv:2406.18655}, 2024.
\newblock \doi{10.48550/arXiv.2406.18655}.

\bibitem[Katzgraber et~al.(2009)Katzgraber, Bomb{\'\i}n, and Martin-Delgado]{katzgraber2009error}
Helmut~G. Katzgraber, H{\'e}ctor Bomb{\'\i}n, and Martin~A. Martin-Delgado.
\newblock Error threshold for color codes and random three-body ising models.
\newblock \emph{Physical Review Letters}, 103:\penalty0 090501, 2009.
\newblock \doi{10.1103/PhysRevLett.103.090501}.

\bibitem[Kesselring et~al.(2018)Kesselring, Pastawski, Eisert, and Brown]{kesselring2018boundaries}
Markus~S. Kesselring, Fernando Pastawski, Jens Eisert, and Benjamin~J. Brown.
\newblock The boundaries and twist defects of the color code and their applications to topological quantum computation.
\newblock \emph{Quantum}, 2:\penalty0 101, 2018.
\newblock \doi{10.22331/q-2018-10-19-101}.

\bibitem[Kesselring et~al.(2024)Kesselring, de~la Fuente, Thomsen, Eisert, Bartlett, and Brown]{kesselring2022anyon}
Markus~S. Kesselring, Julio C.~Magdalena de~la Fuente, Felix Thomsen, Jens Eisert, Stephen~D. Bartlett, and Benjamin~J. Brown.
\newblock Anyon condensation and the color code.
\newblock \emph{PRX Quantum}, 5:\penalty0 010342, 2024.
\newblock \doi{10.1103/PRXQuantum.5.010342}.

\bibitem[Kitaev(1997)]{kitaev1997quantum}
A~Yu Kitaev.
\newblock Quantum computations: algorithms and error correction.
\newblock \emph{Russian Mathematical Surveys}, 52\penalty0 (6):\penalty0 1191, 1997.
\newblock \doi{10.1070/RM1997v052n06ABEH002155}.

\bibitem[Kitaev(2003)]{kitaev2003fault}
A.~Yu Kitaev.
\newblock Fault-tolerant quantum computation by anyons.
\newblock \emph{Annals of Physics}, 303:\penalty0 2--30, 2003.
\newblock \doi{10.1016/S0003-4916(02)00018-0}.

\bibitem[Knill et~al.(1998)Knill, Laflamme, and Zurek]{knill1998resilient}
Emanuel Knill, Raymond Laflamme, and Wojciech~H Zurek.
\newblock Resilient quantum computation: error models and thresholds.
\newblock \emph{Proceedings of the Royal Society of London. Series A: Mathematical, Physical and Engineering Sciences}, 454\penalty0 (1969):\penalty0 365--384, 1998.
\newblock \doi{10.1098/rspa.1998.0166}.

\bibitem[Krinner et~al.(2022)Krinner, Lacroix, Remm, Di~Paolo, Genois, Leroux, Hellings, Lazar, Swiadek, Herrmann, et~al.]{krinner2022realizing}
Sebastian Krinner, Nathan Lacroix, Ants Remm, Agustin Di~Paolo, Elie Genois, Catherine Leroux, Christoph Hellings, Stefania Lazar, Francois Swiadek, Johannes Herrmann, et~al.
\newblock Realizing repeated quantum error correction in a distance-three surface code.
\newblock \emph{Nature}, 605:\penalty0 669--674, 2022.
\newblock \doi{10.1038/s41586-022-04566-8}.

\bibitem[Kschischang et~al.(2001)Kschischang, Frey, and Loeliger]{kschischang2001factor}
Frank~R Kschischang, Brendan~J Frey, and H-A Loeliger.
\newblock Factor graphs and the sum-product algorithm.
\newblock \emph{IEEE Transactions on information theory}, 47:\penalty0 498--519, 2001.
\newblock \doi{10.1109/18.910572}.

\bibitem[Kubica et~al.(2015)Kubica, Yoshida, and Pastawski]{Unfolding}
Aleksander Kubica, Beni Yoshida, and Fernando Pastawski.
\newblock Unfolding the color code.
\newblock \emph{New Journal of Physics}, 17:\penalty0 083026, 2015.
\newblock \doi{10.1088/1367-2630/17/8/083026}.

\bibitem[Kubica(2018)]{kubica2018abcs}
Aleksander~M. Kubica.
\newblock \emph{{The ABCs of the color code: A study of topological quantum codes as toy models for fault-tolerant quantum computation and quantum phases of matter}}.
\newblock PhD thesis, California Institute of Technology, 2018.
\newblock URL \url{https://resolver.caltech.edu/CaltechTHESIS:05282018-173928314}.

\bibitem[Laitinen et~al.(2012)Laitinen, Junttila, and Niemel{\"a}]{laitinen2012extending}
Tero Laitinen, Tommi Junttila, and Ilkka Niemel{\"a}.
\newblock Extending clause learning sat solvers with complete parity reasoning.
\newblock In \emph{2012 IEEE 24th International Conference on Tools with Artificial Intelligence}, volume~1, pages 65--72. IEEE, 2012.
\newblock \doi{10.1109/ICTAI.2012.18}.

\bibitem[Lee et~al.(2024)Lee, Li, and Bartlett]{lee2024color}
Seok-Hyung Lee, Andrew Li, and Stephen~D Bartlett.
\newblock Color code decoder with improved scaling for correcting circuit-level noise.
\newblock \emph{arXiv:2404.07482}, 2024.
\newblock \doi{https://arxiv.org/abs/2404.07482}.

\bibitem[Li(2018)]{li2018fault}
Ying Li.
\newblock Fault-tolerant fermionic quantum computation based on color code.
\newblock \emph{Physical Review A}, 98:\penalty0 012336, 2018.
\newblock \doi{10.1103/PhysRevA.98.012336}.

\bibitem[Maskara et~al.(2019)Maskara, Kubica, and Jochym-O'Connor]{maskara2019advantages}
Nishad Maskara, Aleksander Kubica, and Tomas Jochym-O'Connor.
\newblock Advantages of versatile neural-network decoding for topological codes.
\newblock \emph{Physical Review A}, 99:\penalty0 052351, 2019.
\newblock \doi{10.1103/PhysRevA.99.052351}.

\bibitem[Newman(2022)]{NewmanTalk}
Michael Newman.
\newblock Decoding experimental surface code data.
\newblock Quantum Summer Symposium, 2022.
\newblock URL \url{https://www.youtube.com/watch?v=UP0AoXCT9xU&t=288s&ab_channel=GoogleQuantumAI}.

\bibitem[Nielsen and Chuang(2002)]{nielsen2002quantum}
Michael~A. Nielsen and Isaac Chuang.
\newblock Quantum computation and quantum information, 2002.
\newblock URL \url{10.1017/CBO9780511976667}.

\bibitem[Panteleev and Kalachev(2021)]{panteleev2021degenerate}
Pavel Panteleev and Gleb Kalachev.
\newblock Degenerate quantum ldpc codes with good finite length performance.
\newblock \emph{Quantum}, 5:\penalty0 585, 2021.
\newblock \doi{10.22331/q-2021-11-22-585}.

\bibitem[Postler et~al.(2022)]{postler2022demonstration}
Lukas Postler et~al.
\newblock Demonstration of fault-tolerant universal quantum gate operations.
\newblock \emph{Nature}, 605:\penalty0 675--680, 2022.
\newblock \doi{10.1038/s41586-022-04721-1}.

\bibitem[Preskill(1998)]{preskill1998reliable}
John Preskill.
\newblock Reliable quantum computers.
\newblock \emph{Proceedings of the Royal Society of London. Series A}, 454:\penalty0 385--410, 1998.
\newblock \doi{10.1098/rspa.1998.0167}.

\bibitem[Roffe(2019)]{roffe2019quantum}
Joschka Roffe.
\newblock Quantum error correction: an introductory guide.
\newblock \emph{Contemporary Physics}, 60:\penalty0 226--245, 2019.
\newblock \doi{10.1080/00107514.2019.1667078}.

\bibitem[Roffe(2022)]{Roffe_LDPC_Python_tools_2022}
Joschka Roffe.
\newblock {LDPC: Python tools for low density parity check codes}, 2022.
\newblock URL \url{https://pypi.org/project/ldpc/}.

\bibitem[Roffe et~al.(2020)Roffe, White, Burton, and Campbell]{roffe2020decoding}
Joschka Roffe, David~R White, Simon Burton, and Earl~T. Campbell.
\newblock Decoding across the quantum low-density parity-check code landscape.
\newblock \emph{Physical Review Research}, 2:\penalty0 043423, 2020.
\newblock \doi{10.1103/PhysRevResearch.2.043423}.

\bibitem[Ryan-Anderson et~al.(2021)]{ryan2021realization}
Ryan-Anderson et~al.
\newblock Realization of real-time fault-tolerant quantum error correction.
\newblock \emph{Physical Review X}, 11:\penalty0 041058, 2021.
\newblock \doi{10.1103/PhysRevX.11.041058}.

\bibitem[Sabo et~al.(2024)Sabo, Aloshious, and Brown]{sabo2024trellis}
Eric Sabo, Arun~B Aloshious, and Kenneth~R Brown.
\newblock Trellis decoding for qudit stabilizer codes and its application to qubit topological codes.
\newblock \emph{IEEE Transactions on Quantum Engineering}, 2024.
\newblock \doi{10.1109/TQE.2024.3401857}.

\bibitem[Sahay and Brown(2022)]{sahay2022decoder}
Kaavya Sahay and Benjamin~J. Brown.
\newblock {Decoder for the triangular color code by matching on a M{\"o}bius strip}.
\newblock \emph{PRX Quantum}, 3:\penalty0 010310, 2022.
\newblock \doi{10.1103/PRXQuantum.3.010310}.

\bibitem[San~Miguel et~al.(2023)San~Miguel, Williamson, and Brown]{miguel2022cellular}
Jonathan~F. San~Miguel, Dominic~J. Williamson, and Benjamin~J. Brown.
\newblock A cellular automaton decoder for a noise-bias tailored color code.
\newblock \emph{Quantum}, 7:\penalty0 940, 2023.
\newblock \doi{10.22331/q-2023-03-09-940}.

\bibitem[Sarvepalli and Raussendorf(2012)]{sarvepalli2012efficient}
Pradeep Sarvepalli and Robert Raussendorf.
\newblock Efficient decoding of topological color codes.
\newblock \emph{Physical Review A}, 85:\penalty0 022317, 2012.
\newblock \doi{10.1103/PhysRevA.85.022317}.

\bibitem[Shor(1996)]{shor1996fault}
Peter~W. Shor.
\newblock Fault-tolerant quantum computation.
\newblock In \emph{Proceedings of 37th conference on foundations of computer science}, pages 56--65. IEEE, 1996.
\newblock \doi{10.1109/SFCS.1996.548464}.

\bibitem[Skoric et~al.(2023)Skoric, Browne, Barnes, Gillespie, and Campbell]{skoric2023parallel}
Luka Skoric, Dan~E. Browne, Kenton~M. Barnes, Neil~I. Gillespie, and Earl~T. Campbell.
\newblock Parallel window decoding enables scalable fault tolerant quantum computation.
\newblock \emph{Nature Communications}, 14\penalty0 (1):\penalty0 7040, 2023.
\newblock \doi{10.5281/zenodo.8422904}.

\bibitem[Soos et~al.(2009)Soos, Nohl, and Castelluccia]{soos2009extending}
Mate Soos, Karsten Nohl, and Claude Castelluccia.
\newblock Extending sat solvers to cryptographic problems.
\newblock In \emph{Theory and Applications of Satisfiability Testing-SAT 2009}, pages 244--257. Springer, 2009.
\newblock \doi{10.1007/978-3-642-02777-2_24}.

\bibitem[Steane(1996)]{steane1996error}
Andrew~M. Steane.
\newblock Error correcting codes in quantum theory.
\newblock \emph{Physical Review Letters}, 77:\penalty0 793, 1996.
\newblock \doi{10.1103/PhysRevLett.77.793}.

\bibitem[Stephens(2014)]{stephens2014efficient}
Ashley~M. Stephens.
\newblock Efficient fault-tolerant decoding of topological color codes.
\newblock \emph{arXiv:1402.3037}, 2014.
\newblock \doi{10.48550/arXiv.1402.3037}.

\bibitem[team(2021)]{google2021exponential}
Google~AI team.
\newblock Exponential suppression of bit or phase errors with cyclic error correction.
\newblock \emph{Nature}, 595:\penalty0 383--387, 2021.
\newblock \doi{10.1038/s41586-021-03588-y}.

\bibitem[Thomsen et~al.(2022)Thomsen, Kesselring, Bartlett, and Brown]{thomsen2022low}
Felix Thomsen, Markus~S. Kesselring, Stephen~D. Bartlett, and Benjamin~J. Brown.
\newblock Low-overhead quantum computing with the color code.
\newblock \emph{arXiv:2201.07806}, 2022.
\newblock \doi{10.48550/arXiv.2201.07806}.

\bibitem[Tuckett(2020)]{qecsim}
David~K. Tuckett.
\newblock \emph{Tailoring surface codes: Improvements in quantum error correction with biased noise}.
\newblock PhD thesis, University of Sydney, 2020.
\newblock (qecsim: \url{https://github.com/qecsim/qecsim}).

\bibitem[Tuckett et~al.(2019)Tuckett, Darmawan, Chubb, Bravyi, Bartlett, and Flammia]{tuckett2019tailoring}
David~K. Tuckett, Andrew~S. Darmawan, Christopher~T. Chubb, Sergey Bravyi, Stephen~D. Bartlett, and Steven~T. Flammia.
\newblock Tailoring surface codes for highly biased noise.
\newblock \emph{Phys. Rev. X}, 9\penalty0 (4):\penalty0 041031, 2019.
\newblock \doi{10.1103/PhysRevX.9.041031}.

\bibitem[Wang et~al.(2003)Wang, Harrington, and Preskill]{WANG200331}
Chenyang Wang, Jim Harrington, and John Preskill.
\newblock {Confinement-Higgs transition in a disordered gauge theory and the accuracy threshold for quantum memory}.
\newblock \emph{Annals of Physics}, 303:\penalty0 31--58, 2003.
\newblock \doi{10.1016/S0003-4916(02)00019-2}.

\bibitem[Wille et~al.(2024)]{10646543}
R.~Wille et~al.
\newblock The mqt handbook : A summary of design automation tools and software for quantum computing.
\newblock In \emph{2024 IEEE International Conference on Quantum Software (QSW)}, pages 1--8, 2024.
\newblock \doi{10.1109/QSW62656.2024.00013}.

\end{thebibliography}

\clearpage
\appendix

\section{Three-dimensional \LO{}}\label{sec:app_3d-lo}

In this section, we discuss the \LO{} analogy for the phenomenological noise model, which extends the bit/phase-flip noise model considered in the main text by additionally assuming syndrome noise.

To obtain the parity values of the checks from the encoded quantum state, a syndrome extraction circuit is applied~\cite{roffe2019quantum, nielsen2002quantum}.
Intuitively, for each check, an additional auxiliary qubit that is entangled with the data qubits corresponding to its support is measured to obtain the corresponding $X$ or $Z$ parity value. 
In the bit/phase-flip noise model, we assume that the syndrome extraction circuit is noise-free, i.e., only the data qubits are prone to $X$ or $Z$ Pauli errors. 
In the phenomenological noise model, we additionally assume that the syndrome extraction process is prone to errors, which is why the syndrome measurements need to be repeated.
Phenomenological noise can be modeled as having $X$ or $Z$ errors on data qubits with probability $p$ and classical bit flip errors on measurement outcomes with probability $q$. We assume $p=q$.
Traditionally, the number of rounds of repetition is proportional to the distance of the code~\cite{dennis2002topological}.
Intuitively, the repeated rounds can be thought of as an additional time dimension that is added to the decoding problem, which is why the corresponding decoding graph is three-dimensional.

In Ref.~\cite{gidney2021stim}, a convenient representation of the corresponding syndrome measurement circuit, called the \emph{detector error model} was introduced. 
This model allows us to formulate the decoding problem for more general noise models such as phenomenological noise and circuit-level noise in a similar fashion as the classical decoding problem. 
For an illustrative explanation, we also refer the reader to Ref.~\cite{higgott2023sparse,derks2024designing}.

\begin{figure}[htb]
    \centering
    \includegraphics[scale=0.14]{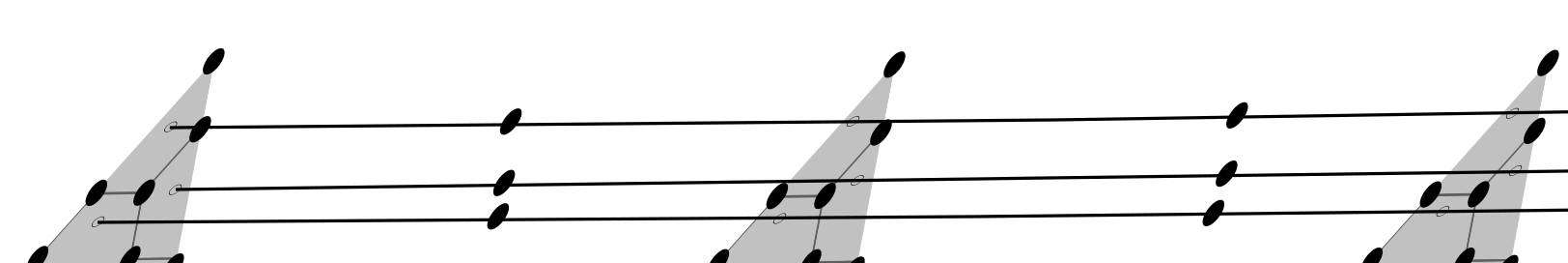}
    \caption{Sketch of the three-dimensional \LO{} puzzle for the 2D color code, corresponding to time-domain decoding of a $d=3$ planar color code under phenomenological noise.
    Each copy of the lattice corresponds to a time step and the additional time-like switches are illustrated as black dots between consecutive layers.
    The incidence between time-like switches and faces is illustrated by circles on the respective faces.\vspace*{-3mm}}
    \label{fig:3d-lo} 
\end{figure}

Intuitively, we can thus indeed model this problem as an instance of a three-dimensional \LO{} puzzle that is constructed from copies of the planar $\LO{}$ lattice, such that there are additional switches between pairs of corresponding faces of consecutive layers that we call \emph{time-like switches}. 
Each layer represents a round of syndrome extraction (i.e., measuring all checks once) and the time-like switches represent measurement errors 
and thus toggle the lights of the two adjacent faces of two consecutive layers in the three-dimensional stack.
This idea is illustrated in~\Cref{fig:3d-lo}.

\section{Additional numerical simulations}
In this section, we present the results of additional numerical simulations.

\subsection{Phenomenological Noise Decoding} 
We have simulated the color code subject to phenomenological noise by constructing Stim~\cite{gidney2021stim} circuits.
We have integrated our decoder with Stim's sampling tool ``sinter''.
Therefore, our decoder can, in principle, be used to decode any stabilizer circuit written as a Stim circuit.
These Stim circuits can be subject to more general noise models than the bit/phase-flip noise model discussed in the main body of the paper.

\begin{figure*}[htb]
     \centering
		\includegraphics[width=\textwidth]{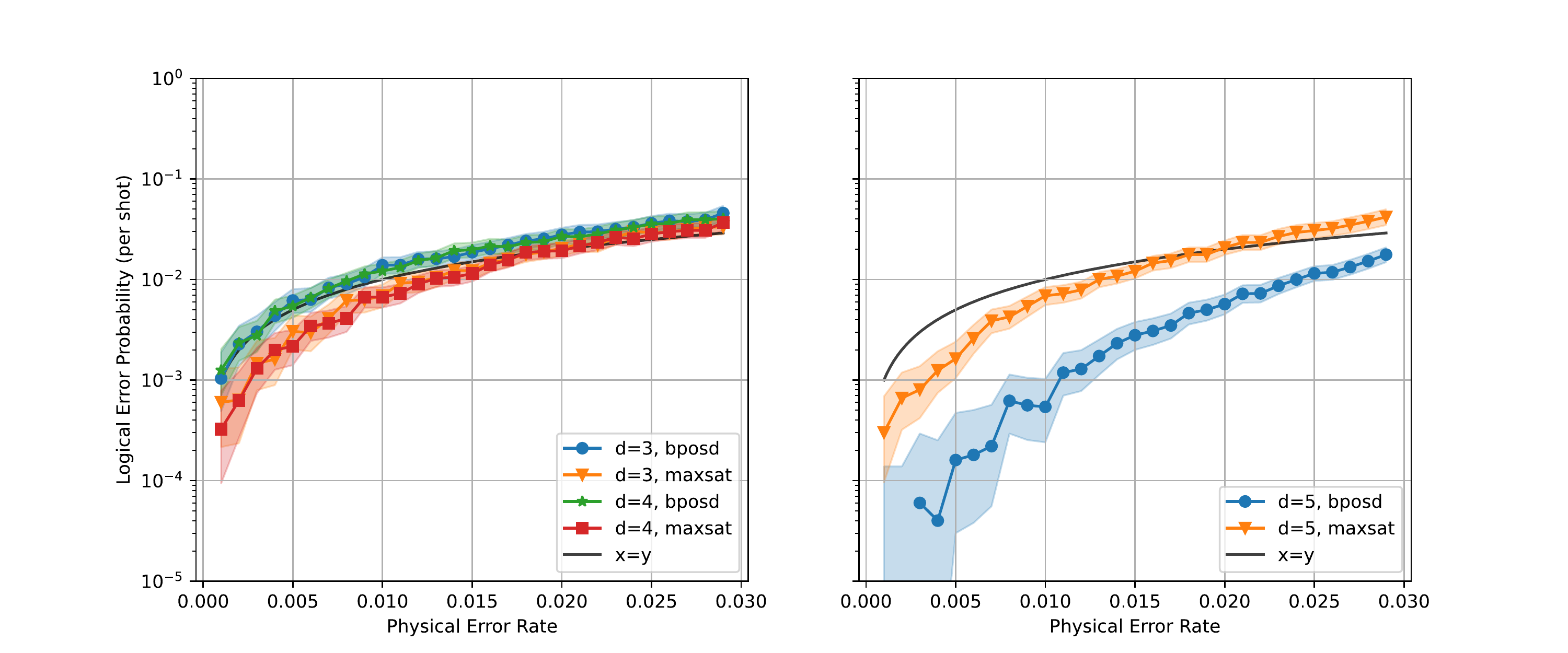}
       \caption{Logical error rate comparison of BP+OSD and the MaxSAT decoder under phenomenological noise for small color codes.
       	Each data point corresponds to collecting up to $10^3$ samples or at most 500 logical errors. 
       	Left: Pseudothreshold---the point at which the logical error rate is equal to the physical error rate---for small color code instances.
       	Right: For larger distances, the decoder performance diverges from BP+OSD.\vspace*{-3mm}}
         \label{fig:pseudoth}
\end{figure*}

We conduct logical error rate simulations for small-distance color codes under phenomenological noise using the proposed MaxSAT decoder.
Moreover, we have used the BP+OSD Stim integration by Higgott~\url{https://github.com/oscarhiggott/stimbposd} to compare the decoding performance of the MaxSAT decoder to the BP+OSD decoder~\cite{roffe2020decoding, panteleev2021degenerate, Roffe_LDPC_Python_tools_2022}, the golden standard decoder for general QLDPC codes.
The results are presented in~\Cref{fig:pseudoth}.

\begin{figure}[htb]
    \centering
    \includegraphics[width=0.9\columnwidth]{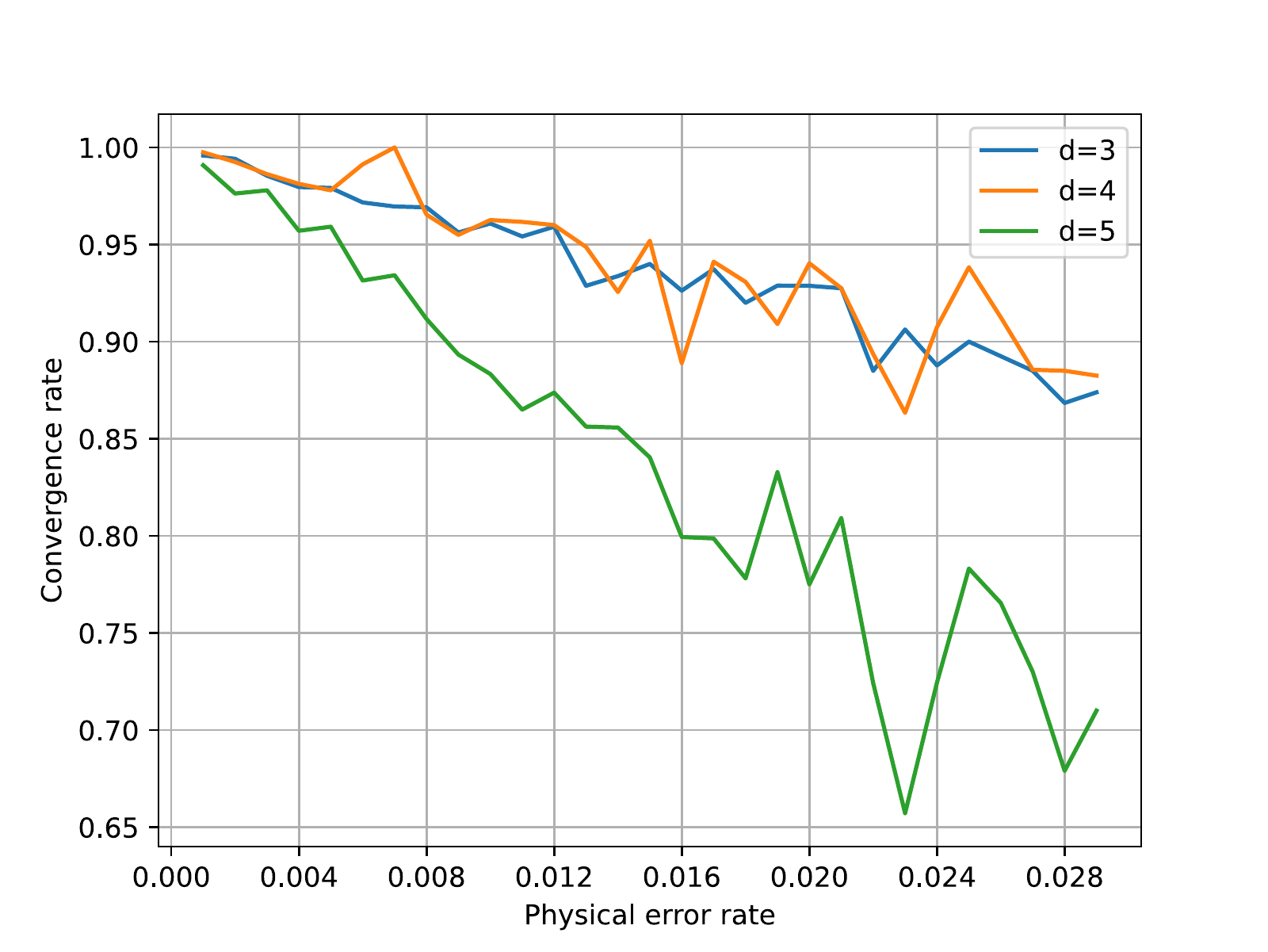}
    \caption{Convergence rate, i.e., the percentage of converged runs of the MaxSAT decoder with timeout \mbox{$t_{max} = 1500ms$} for distance $d\in \{3,4,5\}$ 2D color code instances under phenomenological noise.
    Each data point was obtained from $10^4$ samples and at most 100 logical errors were sampled.\vspace*{-4.5mm}}
    \label{fig:convergence}
\end{figure}

For small distances, we observe that the decoding performance is comparable to BP+OSD and that the decoder exhibits a psuedothreshold. 
However, our simulations also clearly indicate the limitations of the decoder.
In particular, for larger distances, it is not guaranteed that the employed MaxSAT solver converges, i.e., finds the optimal solution within a certain timeout, $t_{max} = 1500ms$. 
In that case, we declare a failure (``non convergance'') and return the all-zeros vector as the recovery operation. 
We observe that for larger distances, this leads to the fact that the decoder (for the considered code and phenomenological noise) does not have a threshold in general.
This aspect is made apparent in ~\Cref{fig:convergence}, which shows that the convergence rate drops significantly for larger distances.

However, there are multiple potential avenues for how this can be amended. 
The most apparent is to make further optimizations of the underlying solver and the satisfiability instance formulation.
We expect that the performance can be boosted considerably by optimizing various aspects on the solver level.
Moreover, in windowed decoding~\cite{skoric2023parallel, dennis2002topological} the size of the decoding problem considered at a time is considerably reduced, making it more amenable for the MaxSAT decoder.
Moreover, also belief propagation (BP), in general, has the problem of not necessarily converging due to loops in the decoding graph. 
For BP this can be amended by using ordered statistics decoding post-processing~\cite{panteleev2021degenerate, roffe2020decoding}.
In a similar spirit, it would be interesting to see how the proposed decoder can be used in combination with another decoder.
We leave such optimizations of the decoder and its application in more general scenarios open for future work.

\subsection{Comparison of different tensor network decoder parameters}\label{sec:app_tn-cmp}
\begin{figure}[htb]
    \centering
    \includegraphics[width=.42\textwidth]{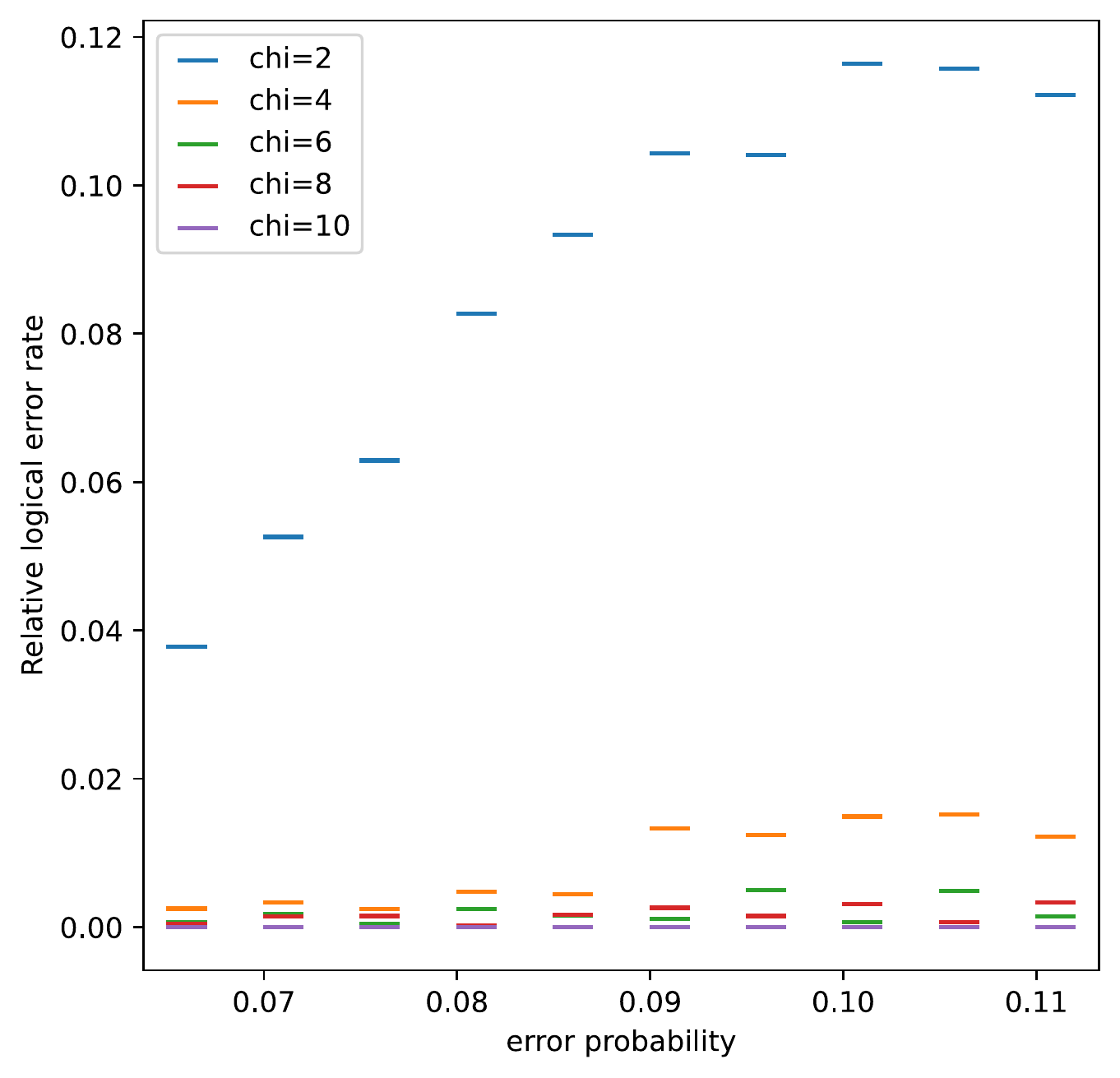}
    \caption{Relative logical error rate $|f_{\chi} - f_{10}|$ for the TN decoder under varying bit/phase-flip noise strength.
    Each data point was obtained from $10^4$ samples.\vspace*{-4.5mm} }
    \label{fig:chi-saturation-tn}
\end{figure}

In the main text, we use the tensor network (TN) decoder~\cite{qecsim} with bond dimension $\chi=6$ (cf.~\Cref{fig:rts-tn}). 
As done previously in related work that uses the TN decoder~\cite{bonilla2021xzzx, tuckett2019tailoring}, we observe that for bond dimension $\chi=6$, the logical error rate essentially saturates.
Hence, further increasing $\chi$ does not yield significant decoding performance improvements (within error bars) as shown in~\Cref{fig:chi-saturation-tn}.

We have furthermore conducted numerical simulations to roughly estimate the runtime performance of the TN decoder for varying bond dimensions.
The results are made publicly available on Github~\url{https://github.com/cda-tum/mqt-qecc} along with all data used in this manuscript.
Finally, we would like to highlight that the investigated runtimes for both the MaxSAT and TN decoder are implementation-specific and, thus, are merely meant to provide a rough frame of reference.
The exact numbers are far from being able to be considered representative or even ``practical''.
The conducted numerical evaluations, however, give a perspective on the MaxSAT decoder's runtime performance relative to the well-established TN decoder.
\end{document}